\documentclass[11pt,a4paper]{article}
\usepackage{jheppub}
\usepackage{amsmath}
\usepackage{amssymb}
\usepackage{amsfonts}
\usepackage{braket}
\usepackage[normalem]{ulem}

\newcommand{\be}{\begin{equation}}
\newcommand{\ee}{\end{equation}}

\title{The Quantum Null Energy Condition, Entanglement Wedge Nesting, and Quantum Focusing}

\author{Chris Akers,}
\author{Venkatesa Chandrasekaran,}
\author{Stefan Leichenauer,}
\author{Adam Levine,}
\author{and Arvin Shahbazi Moghaddam}

\affiliation{Center for Theoretical Physics and Department of Physics,\\
University of California, Berkeley, CA 94720, U.S.A. and}
\affiliation{Lawrence Berkeley National Laboratory, Berkeley, CA 94720, U.S.A.} 
\emailAdd{cakers@berkeley.edu}
\emailAdd{ven\_chandrasekaran@berkeley.edu}
\emailAdd{sleichen@berkeley.edu}
\emailAdd{arlevine@berkeley.edu}
\emailAdd{arvinshm@berkeley.edu}
\abstract{
We study the consequences of Entanglement Wedge Nesting for CFTs with holographic duals. The CFT is formulated on an arbitrary curved background, and we include the effects of curvature-squared couplings in the bulk. In this setup we find necessary and sufficient conditions for Entanglement Wedge Nesting to imply the Quantum Null Energy Condition in $d\leq 5$, extending its earlier holographic proofs. We also show that the Quantum Focusing Conjecture yields the Quantum Null Energy Condition as its nongravitational limit under these same conditions.
}

\begin{document}
\maketitle


\section{Introduction and Summary}


The Quantum Focusing Conjecture (QFC) is a new principle of semiclassical quantum gravity proposed in  \cite{Bousso:2015mna}. Its formulation is motivated by classical focusing, which states that the expansion $\theta$ of a null congruence of geodesics is nonincreasing. Classical focusing is at the heart of several important results of classical gravity~ \cite{Penrose:1964wq,Hawking:1971tu,Hawking:1991nk,Friedman:1993ty}, and likewise quantum focusing can be used to prove quantum generalizations of many of these results \cite{Wall:2009wi,C:2013uza,Bousso:2015eda, Akers:2016aa}.

One of the most important and surprising consequences of the QFC is the Quantum Null Energy Condition (QNEC), which was discovered as a particular nongravitational limit of the QFC~\cite{Bousso:2015mna}. Subsequently the QNEC was proven for free fields~\cite{Bousso:2015wca} and for holographic CFTs on flat backgrounds~\cite{Koeller:2015qmn} (and recently extended in \cite{Fu:2017ab} in a similar way as we do here). The formulation of the QNEC which naturally comes out of the proofs we provide here is as follows.

Consider a codimension-two Cauchy-splitting surface $\Sigma$, which we will refer to as the entangling surface. The Von Neumann entropy $S[\Sigma]$ of the interior (or exterior) or $\Sigma$ is a functional of $\Sigma$, and in particular is a functional of the embedding functions $X^i(y)$ that define $\Sigma$. Choose a one-parameter family of deformed surfaces $\Sigma(\lambda)$, with $\Sigma(0)=\Sigma$, such that (i) $\Sigma(\lambda)$ is given by flowing along null geodesics generated by the null vector field $k^i$ normal to $\Sigma$ for affine time $\lambda$ , and (ii) $\Sigma(\lambda)$ is either ``shrinking" or ``growing" as a function of $\lambda$, in the sense that the domain of dependence of the interior of $\Sigma$ is either shrinking or growing. Then for any point on the entangling surface we can define the combination
\be
T_{ij}(y)k^i(y)k^j(y) - \frac{1}{2\pi}\frac{d}{d\lambda}\left(\frac{k^i(y)}{\sqrt{h(y)}}\frac{\delta S_{\rm ren}}{\delta X^i(y)}\right).
\ee
Here $\sqrt{h(y)}$ is the induced metric determinant on $\Sigma$. Writing this down in a general curved background requires a renormalization scheme both for the energy-momentum tensor $T_{ij}$ and the renormalized entropy $S_{\rm ren}$. Assuming that this quantity is scheme-independent (and hence well-defined), the QNEC states that it is positive. Our main task is to determine the necessary and sufficient conditions we need to impose on $\Sigma$ and the background spacetime at the point $y$ in order that the QNEC hold.

In addition to a proof through the QFC, the holographic proof method of~\cite{Koeller:2015qmn} is easily adaptable to answering this question in full generality. The backbone of that proof is Entanglement Wedge Nesting (EWN), which is a consequence of subregion duality in AdS/CFT~\cite{Akers:2016aa}. A given region on the boundary of AdS is associated with a particular region of the bulk, called the entanglement wedge, which is defined as the bulk region spacelike-related to the extremal surface~\cite{Ryu:2006bv, Hubeny:2007xt,Engelhardt:2014gca,Dong:2017aa} used to compute the CFT entropy on the side toward the boundary region. This bulk region is dual to the given boundary region, in the sense that there is a correspondence between the algebras of operators in the bulk region and the operators in the boundary region which are good semiclassical gravity operators (i.e., they act within the subspace of semiclassical states)~\cite{Czech:2012bh, Jafferis:2015del, Dong:2016eik}. EWN is the statement that nested boundary regions must be dual to nested bulk regions, and clearly follows from the consistency of subregion duality.

While the QNEC can be derived from both the QFC and EWN, there has been no clear connection between these derivations.\footnote{In \cite{Akers:2016aa} it was shown that the QFC in the bulk implies EWN, which in turn implies the QNEC. This is not the same as the connection we are referencing here. The QFC which would imply the boundary QNEC in the sense that we mean is a {\em boundary} QFC, obtained by coupling the boundary theory to gravity.} As it stands, there are apparently two QNECs, the QNEC-from-QFC and the QNEC-from-EWN. We will show in full generality that these two QNECs are in fact the same, at least in $d\leq 5$ dimensions.

Here is a summary of our results:
\begin{itemize}
\item The holographic proof of the QNEC from EWN is extended to CFTs on arbitrary curved backgrounds. In $d=5$ we find necessary that the necessary and sufficient conditions for the ordinary QNEC to hold at a point are that\footnote{Here $\sigma^{(k)}_{ab}$ and $\theta_{(k)}$ are the shear and expansion in the $k^i$ direction, respectively, and $D_a$ is a surface covariant derivative. Our notation is further explained in Appendix~\ref{notation}.}
\be
\theta_{(k)} = \sigma^{(k)}_{ab} = D_a \theta_{(k)} = D_a\sigma^{(k)}_{bc} = R_{ka}= 0
\ee
at that point. For $d<5$ only a subset of these conditions are necessary. This is the subject of \S\ref{state-ind}.

\item We also show holographically that under the weaker set of conditions
\be
\sigma^{(k)}_{ab} = D_a \theta_{(k)} +R_{ka}= D_a\sigma^{(k)}_{bc}= 0
\ee
the Conformal QNEC holds. The Conformal QNEC was introduced in \cite{Koeller:2015qmn} as a conformally-transformed version of the QNEC. This is the strongest inequality that we can get out of EWN. This is the subject of \S\ref{sec:confqnec}

\item By taking the non-gravitational limit of the QFC we are able to derive the QNEC again under the same set of conditions as we did for EWN. This is the subject of \S\ref{sec:qnecqfc}.

\item We argue in \S\ref{schemeind} that the statement of the QNEC is scheme-independent whenever the conditions that allow us to prove it hold. This shows that the two proofs of the QNEC are actually proving the same, unambiguous field--theoretic bound.
 
\end{itemize}

We conclude in \S\ref{discussion} with a discussion and suggest future directions. A number of technical Appendices are included as part of our analysis.

\paragraph{Relation to other work}

While this work was in preparation, \cite{Fu:2017ab} appeared which has overlap with our discussion of EWN and the scheme-independence of the QNEC. The results of \cite{Fu:2017ab} relied on a number of assumptions about the background: the null curvature condition and a positive energy condition. From this they derive certain sufficient conditions for the QNEC to hold. We do not assume anything about our backgrounds a priori, and include all relevant higher curvature corrections. This gives our results greater generality, as we are able to find both necessary and sufficient conditions for the QNEC to hold.


\section{Entanglement Wedge Nesting}

\subsection{Subregion Duality}\label{sec-SubregionDuality}

The statement of AdS/CFT includes a correspondence between operators in the semiclassical bulk gravitational theory and CFT operators on the boundary. Moreover, it has been shown~\cite{Harlow:2016vwg,Dong:2016eik} that such a correspondence exists between the operator algebras of subregions in the CFT and certain associated subregions in the bulk as follows: Consider a spatial subregion $A$ in the boundary geometry. The extremal surface anchored to $\partial A$, which is used to compute the entropy of $A$~\cite{Ryu:2006bv, Hubeny:2007xt}, bounds the so-called entanglement wedge of $A$, $\mathcal{E}(A)$, in the bulk. More precisely $\mathcal{E}(A)$ is the codimension-zero bulk region spacelike-related to the extremal surface on the same side of the extremal surface as $A$. Subregion duality is the statement that the operator algebras of $\mathcal{D}(A)$ and $\mathcal{E}(A)$ are dual, where $\mathcal{D}(A)$ denotes the domain of dependence of $A$.

\paragraph{Entanglement Wedge Nesting} 
The results of this section follow from EWN, which we now describe. Consider two boundary regions $A_1$ and $A_2$ such that $\mathcal{D}(A_{1})\subseteq \mathcal{D}(A_{2})$. Then consistency of subregion duality implies that $\mathcal{E}(A_{1})\subseteq \mathcal{E}(A_{2})$ as well, and this is the statement of EWN. In particular, EWN implies that the extremal surfaces associated to $A_1$ and $A_2$ cannot be timelike-related.

\begin{figure}
	\centering
	\includegraphics[scale=.7]{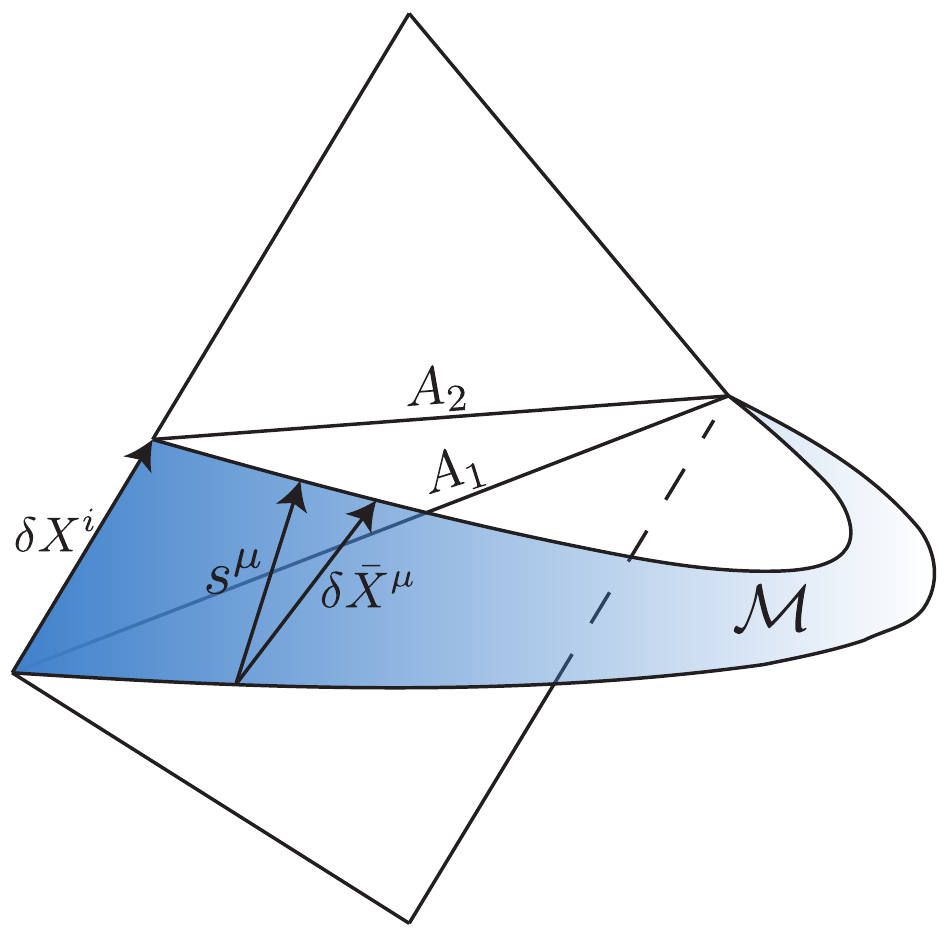}
	\caption{Here we show the holographic setup which illustrates Entanglement Wedge Nesting. A spatial region $A_1$ on the boundary is deformed into the spatial region $A_2$ by the null vector $\delta X^i$. The extremal surfaces of $A_1$ and $A_2$ are connected by a codimension-one bulk surface $\mathcal{M}$ (shaded blue) that is nowhere timelike by EWN. Then the vectors $\delta \bar{X}^\mu$ and $s^\mu$, which lie in $\mathcal{M}$, have nonnegative norm.}\label{EWNfig}
\end{figure}

We will mainly be applying EWN to the case of a one-paramter family of boundary regions, $A(\lambda)$, where $\mathcal{D}(A(\lambda_{1}))\subseteq \mathcal{D}(A(\lambda_{2}))$ whenever $\lambda_1 \leq \lambda_2$. Then the union of the one-parameter family of extremal surfaces associated to $A(\lambda)$ forms a codimension-one surface in the bulk that is nowhere timelike. We denote this codimension-one surface by $\mathcal{M}$. See Fig.~\ref{EWNfig} for a picture of the setup.

Since $\mathcal{M}$ is nowhere timelike, every one of its tangent vectors must have nonnegative norm. In particular, consider the embedding functions $\bar{X}^\mu$ of the extremal surfaces in some coordinate system. Then the vectors $\delta \bar{X}^\mu \equiv \partial_\lambda \bar{X}^\mu$ is tangent to $\mathcal{M}$, and represents a vector that points from one extremal surface to another. Hence we have $(\delta \bar{X})^2 \geq 0$ from EWN, and this is the inequality that we will discuss for most of the remainder of this section.

Before moving on, we will note that $(\delta \bar{X})^2 \geq 0$ is not necessarily the strongest inequality we get from EWN. At each point on $\mathcal{M}$, the vectors which are tangent to the extremal surface passing through that point are known to be spacelike. Therefore if $\delta \bar{X}^\mu$ contains any components which are tangent to the extremal surface, they will serve to make the inequality $(\delta \bar{X})^2 \geq 0$ weaker. We define the vector $s^\mu$ at any point of $\mathcal{M}$ to be the part of $\delta \bar{X}^\mu$ orthogonal to the extremal surface passing through that point. Then $(\delta \bar{X})^2 \geq s^2\geq 0$. We will discuss the $s^2 \geq 0$ inequality in \S\ref{sec:confqnec} after handling the $(\delta \bar{X})^2 \geq 0$ case.

\subsection{Near-Boundary EWN}

In this section we explain how to calculate the vector $\delta \bar{X}^\mu$ and $s^\mu$ near the boundary explicitly in terms of CFT data. Then the EWN inequalities $(\delta \bar{X})^2 >0$ and $s^2>0$ can be given a CFT meaning. The strategy is to use a Fefferman-Graham expansion of both the metric and extremal surface, leading to equations for $\delta \bar{X}^\mu$ and $s^\mu$ as power series in the bulk coordinate $z$ (including possible log terms). In the following sections we will analyze the inequalities that are derived in this section.

\paragraph{Bulk Metric}
We work with a bulk theory in $AdS_{d+1}$ that consists of Einstein gravity plus curvature-squared corrections. For $d\leq 5$ this is the complete set of higher curvature corrections that have an impact on our analysis. The Lagrangian is\footnote{For simplicity we will not include matter fields explicitly in the bulk, but their presence should not alter any of our conclusions.}
\be
\mathcal{L} = \frac{1}{16\pi G_N} \left(\frac{d(d-1)}{\tilde L^2} + \mathcal{R} + \ell^2 \lambda_1 \mathcal{R}^2 + \ell^2 \lambda_2 \mathcal{R}_{\mu\nu}^2+ \ell^2 \lambda_{\rm GB}\mathcal{L}_{\rm GB} \right),
\ee
where $\mathcal{L}_{GB} = \mathcal{R}_{\mu\nu\rho\sigma}^2 -4 \mathcal{R}_{\mu\nu}^2 + \mathcal{R}^2$ is the Gauss--Bonnet Lagrangian, $\ell^2$ is the cutoff scale, and $\tilde L^2$ is the scale of the cosmological constant. The bulk metric has the following near boundary expansion in Fefferman-Graham gauge \cite{deHaro:2000xn}:
\begin{align}
ds^2 &=\frac{L^2}{z^2}(dz^2+\bar{g}_{ij}(x,z)dx^{i}dx^{j}),\\
\bar{g}_{ij}(x,z) &= g^{(0)}_{ij}(x)+z^2 g^{(2)}_{ij}(x)+z^4 g^{(4)}_{ij}(x)+\ldots+z^d\log z\, g^{(d,{\rm log})}_{ij}(x) + z^d g^{(d)}_{ij}(x)+o(z^d).\label{eq-gexp}
\end{align}
Note that the length scale $L$ is different from $\tilde L$, but the relationship between them will not be important for us. Demanding that the above metric solve bulk gravitational equations of motion gives expressions for all of the $g^{(n)}_{ij}$ for $n<d$, including $g^{(d,{\rm log})}_{ij}(x)$, in terms of $g^{(0)}_{ij}(x)$. This means, in particular, that these terms are all state-independent. One finds that $g^{(d,{\rm log})}_{ij}(x)$ vanishes unless $d$ is even. We provide explicit expressions for some of these terms in Appendix \ref{zexp}.

The only state-dependent term we have displayed, $g^{(d)}_{ij}(x)$, contains information about the expectation value of the energy-momentum tensor $T_{ij}$ of the field theory. In odd dimensions we have the simple formula~\cite{Faulkner:2013ica}\footnote{Even though \cite{Faulkner:2013ica} worked with a flat boundary theory, one can check that this formula remains unchanged when the boundary is curved.}
\be\label{eq-gd}
g^{(d={\rm odd})}_{ij}=\frac{16\pi G_{N}}{\eta d L^{d-1}}\langle T_{ij}\rangle,
\ee
with 
\be
\eta = 1-2\left(d(d+1)\lambda_1+d\lambda_2+(d-2)(d-3)\lambda_{\rm GB}\right)\frac{\ell^2}{L^2}
\ee
In even dimensions the formula is more complicated. For $d=4$ we discuss the form of the metric in Appendix~\ref{d=4}

\paragraph{Extremal Surface}
EWN is a statement about the causal relation between entanglement wedges. To study this, we need to calculate the position of the extremal surface. We parametrize our extremal surface by the coordinate $(y^a,z)$, and the position of the surface is determined by the embedding functions $\bar{X}^{\mu}(y^a,z)$. The intrinsic metric of the extremal surface is denoted by $\bar{h}_{\alpha\beta}$, where $\alpha =(a,z)$. For convenience we will impose the gauge conditions $\bar{X}^z = z$ and $\bar{h}_{az} =0$.

The functions $\bar{X}(y^a,z)$ are determined by extremizing the generalized entropy~\cite{Engelhardt:2014gca, Dong:2017aa} of the entanglement wedge. This generalized entropy consists of geometric terms integrated over the surface as well as bulk entropy terms. We defer a discussion of the bulk entropy terms to \S\ref{sec-bulkent} and write only the geometric terms, which are determined by the bulk action:
\begin{align}
S_{\rm gen} = \frac{1}{4G_N} \int \sqrt{\bar{h}}\left[1 + 2\lambda_1\ell^2 \mathcal{R} + \lambda_2\ell^2\left(\mathcal{R}_{\mu\nu}\mathcal{N}^{\mu\nu} - \frac{1}{2}\mathcal{K}_\mu\mathcal{K}^\mu\right) + 2\lambda_{\rm GB}\ell^2\bar{r}\right].
\end{align}
We discuss this entropy functional in more detail in Appendix~\ref{ExSurf}. The Euler-Lagrange equations for $S_{\rm gen}$ are the equations of motion for $\bar{X}^\mu$. Like the bulk metric, the extremal surface equations can be solved at small-$z$ with a Fefferman--Graham-like expansion:
\begin{align}\label{eq-Xexp}
\bar{X}^{i}(y,z)=X_{(0)}^{i}(y)+z^2 X_{(2)}^{i}(y)+z^4 X_{(4)}^{i}(y)+\ldots+z^d\log z\, X^i_{(d,{\rm log})}(y) +z^d X_{(d)}^{i}(y)+o(z^d),
\end{align}
As with the metric, the coefficient functions $X^i_{(n)}$ for $n<d$, including the log term, can be solved for in terms of $X_{(0)}^i$ and $g^{(0)}_{ij}$, and again the log term vanishes unless $d$ is even. The state-dependent term $X_{(d)}^{i}$ contains information about variations of the CFT entropy, as we explain below.

\paragraph{The $z$-Expansion of EWN} 
By taking the derivative of \eqref{eq-Xexp} with respect to $\lambda$, we find the $z$-expansion of $\delta \bar{X}^i$. We will discuss how to take those derivatives momentarily. But given the $z$-expansion of $\delta \bar{X}^i$, we can combine this with the $z$-expansion of $\bar{g}_{ij}$ in \eqref{eq-gexp} to get the $z$-expansion of $(\delta \bar{X})^2$:
\begin{align}\label{eq-deltaXexp}
\frac{z^2}{L^2}(\delta \bar{X})^2 = g_{ij}^{(0)} \delta X_{(0)}^i \delta X_{(0)}^j + z^2\left(2g_{ij}^{(0)} \delta X_{(0)}^i \delta X_{(2)}^j +g_{ij}^{(2)} \delta X_{(0)}^i \delta X_{(0)}^j+X_{(2)}^m\partial_m g_{ij}^{(0)} \delta X_{(0)}^i \delta X_{(0)}^j \right) +\cdots
\end{align}
EWN implies that $(\delta \bar{X})^2 \geq0$, and we will spend the next few sections examining this inequality using the expansion  \eqref{eq-deltaXexp}. From the general arguments given above, we can get a stronger inequality by considering the vector $s^\mu$  and its norm rather than $\delta \bar{X}^\mu$. The construction of $s^\mu$ is more involved, but we would similarly construct an equation for $s^2$ at small $z$. We defer further discussion of $s^\mu$ to \S\ref{sec:confqnec}.

Now we return to the question of calculating $\delta \bar{X}^i$. Since all of the $X_{(n)}^i$ for $n<d$ are known explicitly from solving the equation of motion, the $\lambda$-derivatives of those terms can be taken and the results expressed in terms of the boundary conditions for the extremal surface. The variation of the state-dependent term, $\delta X_{(d)}^i$, is also determined by the boundary conditions in principle, but in a horribly non-local way. However, we will now show that $X_{(d)}^i$ (and hence $\delta X_{(d)}^i$) can be re-expressed in terms of variations of the CFT entropy.

\paragraph{Variations of the Entropy}
The CFT entropy $S_{\rm CFT}$ is equal to the generalized entropy $S_{\rm gen}$ of the entanglement wedge in the bulk. To be precise, we need to introduce a cutoff at $z = \epsilon$ and use holographic renormalization to properly define the entropy. Then we can use the calculus of variations to determine variations of the entropy with respect to the boundary conditions at $z=\epsilon$. There will be terms which diverge as $\epsilon\to 0$, as well as a {\rm finite} term, which is the only one we are interested in at the moment. In odd dimensions, the finite term is given by a simple integral over the entangling surface in the CFT:
\be\label{eq-Svar}
\left. \delta S_{\rm CFT} \right|_{\rm finite} = \eta d L^{d-1} \int d^{d-2}y \sqrt{h} g_{ij}X_{(d)}^i\delta X^j.
\ee 
This finite part of $S_{\rm CFT}$ is the renormalized entropy, $S_{\rm ren}$, in holographic renormalization. Eventually we will want to assure ourselves that our results are scheme-independent. This question was studied in~\cite{Fu:2017aa}, and we will discuss it further in \S\ref{schemeind}. For now, the important take-away from \eqref{eq-Svar} is
\be\label{eq-Xd}
\frac{1}{\sqrt{h}}\frac{\delta S_{\rm ren}}{\delta X^i(y)} = - \frac{\eta dL^{d-1}}{4G_N}X_{(d,{\rm odd})}^i.
\ee
The case of even $d$ is more complicated, and we will cover the $d=4$ case in Appendix~\ref{d=4}.

\subsection{State-Independent Inequalities \label{state-ind}}
The basic EWN inequality is $(\delta \bar{X})^2 \geq 0$. The challenge is to write this in terms of boundary quantities. In this section we will look at the state-independent terms in the expansion of \eqref{eq-deltaXexp}. The boundary conditions at $z=0$ are given by the CFT entangling surface and background geometry, which we denote by $X^i$ and $g_{ij}$ without a $(0)$ subscript. The variation vector of the entangling surface is the null vector $k^i = \delta X^i$. We can use the formulas of Appendix~\ref{EWN} to express the other $X_{(n)}^i$ for $n<d$ in terms of $X^i$ and $g_{ij}$. This allows us to express the state-independent parts of $(\delta \bar{X})^2 \geq 0$ in terms of CFT data. In this subsection we will look at the leading and subleading state-independent parts. These will be sufficient to fully cover the cases $d\leq 5$.

\paragraph{Leading Inequality}

From \eqref{eq-deltaXexp}, we see that the first term is actually $k_ik^i =0$. The next term is the one we call the leading term, which is
\be
\left.L^{-2}(\delta \bar{X})^2\right|_{z^0} = 2k_i \delta X^i_{(2)} + g_{ij}^{(2)}k^ik^j+X_{(2)}^m\partial_m g_{ij}k^ik^j.
\ee
From \eqref{eq-X2}, we easily see that this is equivalent to
\be
L^{-2}\left.(\delta \bar{X}^i)^2 \right|_{z^0} = \frac{1}{(d-2)^2}\theta_{(k)}^2 + \frac{1}{d-2}\sigma_{(k)}^2,
\ee
where $\sigma^{(k)}_{ab}$ and $\theta_{(k)}$ are the shear and expansion of the null congruence generated by $k^i$, and are given by the trace and trace-free parts of $k_iK^i_{ab}$, with $K^i_{ab}$ the extrinsic curvature of the entangling surface. This leading inequality is always nonnegative, as required by EWN. Since we are in the small-$z$ limit, the subleading inequality is only relevant when this leading inequality is saturated. So in our analysis below we will focus on the $\theta_{(k)} = \sigma^{(k)}_{ab}= 0 $ case, which can always be achieved by choosing the entangling surface appropriately. Note that in $d=3$ this is the only state-independent term in $(\delta \bar{X})^2$, and furthermore we always have $\sigma^{(k)}_{ab}=0$ in $d=3$.

\paragraph{Subleading Inequality}

The subleading term in $(\delta \bar{X})^2$ is order $z^2$ in $d\geq5$, and order $z^2\log z$ in $d=4$. These two cases are similar, but it will be easiest to focus first on $d\geq5$ and then explain what changes in $d=4$. The terms we are looking for are
\begin{align}
\left.L^{-2}(\delta \bar{X})^2\right|_{z^2} &= 2k_i \delta X^i_{(4)} + 2g_{ij}^{(2)}k^i\delta X_{(2)}^j+g_{ij}\delta X_{(2)}^i\delta X_{(2)}^j+g_{ij}^{(4)}k^ik^j+X^m_{(4)}\partial_mg_{ij}k^ik^j\nonumber\\
&+2X_{(2)}^m\partial_mg_{ij}k^i\delta X_{(2)}^j+X^m_{(2)}\partial_mg_{ij}^{(2)}k^ik^j+\frac{1}{2}X_{(2)}^mX_{(2)}^n\partial_m\partial_ng_{ij}k^ik^j.\label{eq-subleadinginitial}
\end{align}
This inequality is significantly more complicated than the previous one. The details of its evaluation are left to Appendix~\ref{EWN}. The result, assuming $\theta_{(k)} = \sigma^{(k)}_{ab} = 0$, is
\begin{align}
\left.L^{-2}(\delta \bar{X})^2\right|_{z^2}  &=\frac{1}{4(d-2)^2}(D_a\theta_{(k)} +2R_{ka})^2\nonumber\\
&+\frac{1}{(d-2)^2(d-4)}(D_a\theta_{(k)} + R_{ka})^2+\frac{1}{2(d-2)(d-4)}(D_a\sigma^{(k)}_{bc})^2 \nonumber\\
&+\frac{\kappa}{d-4}\left(C_{kabc}C_{k}^{~abc}-2C_{k~ca}^{~c}C_{k~b}^{~b~a}\right).
 \label{EWN Subleading}
\end{align}
where $\kappa$ is proportional to $\lambda_{\rm GB} \ell^2/L^2$ and is defined in Appendix~\ref{EWN}. Aside from the Gauss--Bonnet term we have a sum of squares, which is good because EWN requires this to be positive when $\theta_{(k)}$ and $\sigma_{(k)}$ vanish. Since $\kappa \ll 1$, it cannot possibly interfere with positivity unless the other terms were zero. This would require $D_a\theta_{(k)} = D_a\sigma^{(k)}_{bc} = R_{ka} = 0$ in addition to our other conditions. But, following the arguments of~\cite{Leichenauer:2017bmc}, this cannot happen unless the components $C_{kabc}$ of the Weyl tensor also vanish at the point in question. Thus EWN is always satisfied. Also noteSecond, the last two terms in middle line of \eqref{EWN Subleading} are each conformally invariant when $\theta_{(k)} = \sigma^{(k)}_{ab} = 0$, which we have assumed. This will become important later.

Finally, though we have assumed $d \geq 5$ to arrive at this result, we can use it to derive the expression for $\left.L^{-2}(\delta \bar{X})^2\right|_{z^2\log z}$ in $d=4$. The rule, explained in Appendix~\ref{d=4}, is to multiply the RHS by $4-d$ and then set $d=4$. This has the effect of killing the conformally non-invariant term, leaving us with 
\begin{align}
\left.L^{-2}(\delta \bar{X})^2\right|_{z^2\log z, d=4}  &=-\frac{1}{4}(D_a\theta_{(k)} + R_{ka})^2-\frac{1}{4}(D_a\sigma^{(k)}_{bc})^2.
\end{align}
The Gauss--Bonnet term also disappears because of a special Weyl tensor identity in $d=4$~\cite{Fu:2017aa}. The overall minus sign is required since $\log z <0$ in the small $z$ limit. In addition, we no longer require that $R_{ka}$ and $D_a\theta_{(k)}$ vanish individually to saturate the inequality: only their sum has to vanish. This still requires that $C_{kabc} =0$, though.


\subsection{The Quantum Null Energy Condition }

The previous section dealt with the two leading state-independent inequalities that EWN implies. Here we deal with the leading state-\emph{dependent} inequality, which turns out to be the QNEC. 

At all orders lower than $z^{d-2}$, $(\delta \bar{X})^2$ is purely geometric. At order $z^{d-2}$, however, the CFT energy-momentum tensor enters via the Fefferman--Graham expansion of the metric, and variations of the entropy enter through $X_{(d)}^i$. In odd dimensions the analysis is simple and we will present it here, while in general even dimensions it is quite complicated. Since our state-independent analysis is incomplete for $d>5$ anyway, we will be content with analyzing only $d=4$ for the even case. The $d=4$ calculation is presented in Appendix~\ref{d=4}. Though is it more involved that the odd-dimensional case, the final result is the same.

Consider first the case where $d$ is odd. Then we have
\be\label{eq:ewnd}
\left.L^{-2}(\delta \bar{X})^2\right|_{z^{d-2}} = g_{ij}^{(d)}k^ik^j + 2k_i\delta X^i_{(d)} + X_{(d)}^m\partial_m g_{ij} k^ik^j= g_{ij}^{(d)}k^ik^j + 2\delta\left(k_i\delta X^i_{(d)}\right).
\ee
From \eqref{eq-gd} and \eqref{eq-Xd}, we find that 
\be
\left.L^{-2}(\delta \bar{X})^2\right|_{z^{d-2}} = \frac{16\pi G_{N}}{\eta d L^{d-1}}\left[\langle T_{kk}\rangle- \delta\left(\frac{k^i}{2\pi\sqrt{h}}\frac{\delta S_{\rm ren}}{\delta X^i}\right)\right].\label{eq-EWNQNEC}
\ee
The nonnegativity of the term in brackets is equivalent to the QNEC. The case where $d$ is even is more complicated, and we will go over the $d=4$ case in Appendix~\ref{d=4}.


\subsection{The Conformal QNEC}\label{sec:confqnec}

As mentioned in \S\ref{sec-SubregionDuality}, we can get a stronger inequality from EWN by considering the norm of the vector $s^\mu$, which is the part of $\delta \bar{X}^\mu$ orthogonal to the extremal surface. Our gauge choice $\bar{X}^z=z$ means that $s^\mu \neq \delta \bar{X}^\mu$, and so we get a nontrivial improvement by considering $s^2 \geq 0$ instead of $(\delta \bar{X})^2 \geq 0$.

We can actually use the results already derived above to compute $s^2$ with the following trick. We would have had $\delta \bar{X}^\mu = s^\mu$ if the surfaces of constant $z$ were already orthogonal to the extremal surfaces. But we can change our definition of the constant-$z$ surfaces with a coordinate transformation in the bulk to make this the case, apply the above results to $(\delta \bar{X})^2$ in the new coordinate system, and then transform back to the original coordinates. The coordinate transformation we are interested in performing is a PBH transformation~\cite{Imbimbo:1999bj}, since it leaves the metric in Fefferman--Graham form, and so induces a Weyl transformation on the boundary.

So from the field theory point of view, we will just be calculating the consequences of EWN in a different conformal frame, which is fine because we are working with a CFT. With that in mind it is easy to guess the outcome: the best conformal frame to pick is one in which all of the non-conformally-invariant parts of the state-independent terms in $(\delta \bar{X})^2$ are set to zero, and when we transform the state-dependent term in the new frame back to the original frame we get the so-called Conformal QNEC first defined in~\cite{Koeller:2015qmn}. This is indeed what happens, as we will now see.

\paragraph{Orthogonality Conditions}

First, we will examine in detail the conditions necessary for $\delta \bar{X}^\mu = s^\mu$, and their consequences on the inequalities derived above. We must check that 
\be
\bar{g}_{ij} \partial_\alpha \bar{X}^i \delta \bar{X}^j = 0.
\ee
for both $\alpha =z$ and $\alpha = a$. As above, we will expand these conditions in $z$. When $\alpha =z$, at lowest order in $z$ we find the condition
\be
0 = k_iX^i_{(2)} ,
\ee
which is equivalent to $\theta_{(k)} = 0$. When $\alpha = a$, the lowest-order in $z$ inequality is automatically satisfied because $k^i$ is defined to be orthogonal to the entangling surface on the boundary. But at next-to-lowest order we find the condition
\begin{align}
0 &= k_i\partial_a X_{(2)}^i + e_{ai}\delta X_{(2)}^i + g_{ij}^{(2)}e_a^ik^j+ X_{(2)}^m\partial_m g_{ij}e_a^ik^j\\
&= -\frac{1}{2(d-2)}\left[(D_a-2w_a)\theta_{(k)} + 2R_{ka}\right].
\end{align}
Combined with the $\theta_{(k)}=0$ condition, this tells us that that $D_a\theta_{(k)} = -2R_{ka}$ is required. When these conditions are satisfied, the state-dependent terms of $(\delta \bar{X})^2$ analyzed above become\footnote{We have not included some terms at order $z^2$ which are proportional to $\sigma^{(k)}_{ab}$ because they never play a role in the EWN inequalities.}
\begin{align}\label{eq:deltax}
L^{-2}(\delta \bar{X})^2 &= \frac{1}{d-2} \sigma_{(k)}^2+ \left[\frac{1}{(d-2)^2(d-4)}(R_{ka})^2+\frac{1}{2(d-2)(d-4)}(D_a\sigma^{(k)}_{bc})^2\right]z^2 + \cdots 
\end{align}
Next we will demonstrate that $\theta_{(k)}=0$ and $D_a\theta_{(k)} = -2R_{ka}$ can be achieved by a Weyl transformation, and then use that fact to write down the $s^2\geq 0$ inequality that we are after.

\paragraph{Achieving $\delta \bar{X}^\mu = s^\mu$ with a Weyl Transformation}

Our goal now is to begin with a generic situation in which $\delta \bar{X}^\mu \neq s^\mu$ and use a Weyl transformation to set $\delta \bar{X}^\mu \to s^\mu$. This means finding a new conformal frame with $\hat{g}_{ij} = e^{2\phi(x)} g_{ij}$ such that $\hat{\theta}_{(k)}=0$ and $\hat{D}_a\hat{\theta}_{(k)} = -2\hat{R}_{ka}$, which would then imply that $\delta \hat{X}^\mu =s^\mu$ (we omit the bar on $\delta \hat{X}^\mu$ to avoid cluttering the notation, but logically it would be $\delta \hat{\bar{X}}^\mu$).

Computing the transformation properties of the geometric quantities involved is a standard exercise, but there is one extra twist involved here compared to the usual prescription. Ordinarily a vector such as $k^i$ would be invariant under the Weyl transformation. However, for our setup is it is important that $k^i$ generate an affine-parameterized null geodesic. Even though the null geodesic itself is invariant under Weyl transofrmation, $k^i$ will no longer be the correct generator. Instead, we have to use $\hat{k}^i = e^{-2\phi}k^i$. Another way of saying this is that $k_i = \hat{k}_i$ is invariant under the Weyl transformation. With this in mind, we have
\begin{align}
e^{2 \phi} \hat{R}_{ka}  &= R_{ka}- (d-2) \left[ D_a \partial_k \phi - w_a \partial_k \phi - k_j K^j_{ab}\partial^b \phi - \partial_k \phi \partial_a \phi \right],\\
e^{2 \phi} \hat{\theta}_{(k)} &=\theta_{(k)} + (d-2) \partial_k \phi,\\
e^{2\phi}\hat{D}_a \hat{\theta}_{(k)} &=D_a \theta_{(k)} + (d-2) D_a \partial_k \phi - 2 \theta_{(k)}\partial_a \phi  - 2(d-2) \partial_k \phi\partial_a \phi,\\
\hat{\sigma}^{(k)}_{ab} &= \sigma^{(k)}_{ab}, \\
\hat{D}_c \hat{\sigma}^{(k)}_{ab} &= D_c \sigma^{(k)}_{ab} - 2\left[\sigma^{(k)}_{c(b} \partial_{a)} \phi + \sigma^{(k)}_{ab} \partial_c \phi - g_{c(a} \sigma^{(k)}_{b)d} \nabla^d \phi\right],\\
\hat{w}_a &= w_a - \partial_a \phi.
\end{align}
So we may arrange $\hat{\theta}_{(k)} = 0$ at a given point on the entangling surface by choosing $\partial_k \phi = -\theta_{(k)}/(d-2)$ that that point. Having chosen that, and assuming $\sigma_{ab}^{(k)}$ =0 at the same point, one can check that
\be
e^{2\phi}\left(\hat{D}_a \hat{\theta}_{(k)} + 2\hat{R}_{ka}\right) =  D_a \theta_{(k)} -2w_a\theta_{(k)} +2R_{ka}- (d-2)D_a \partial_k \phi
\ee
So we can choose $D_a \partial_k \phi$ to make the combination $\hat{D}_a \hat{\theta}_{(k)} + 2\hat{R}_{ka}$ vanish. Then in the new frame we have $\delta \hat{X}^\mu =s^\mu$.

\paragraph{The $s^2\geq$ Inequality}
Based on the discussion above, we were able to find a conformal frame that allows us to compute the $s^2$. For the state-independent parts we have
\begin{align}
L^{-2} s^2 &= \frac{1}{d-2} \hat\sigma_{(k)}^2+ \left[\frac{1}{(d-2)^2(d-4)}(\hat R_{ka})^2+\frac{1}{2(d-2)(d-4)}(\hat D_a\hat \sigma^{(k)}_{bc})^2\right]\hat{z}^2 + \cdots 
\end{align}
Here we also have a new bulk coordinate $\hat{z} = ze^\phi$ associated with the bulk PBH transformation. All we have to do now is transform back into the original frame to find $s^2$. Since $\hat{\theta}_{(k)} = \hat{D}_a \hat{\theta}_{(k)} + 2\hat{R}_{ka}=0$, we actually have that
\be
\hat R_{ka} = \hat{D}_a\hat \theta_{(k)} - \hat{w}_a \hat \theta_{(k)} - \hat{R}_{ka},
\ee
which transforms homogeneously under Weyl transformations when $\sigma^{(k)}_{ab} =0$. Thus, up to an overall scaling factor, we have
\begin{align}
L^{-2} s^2 &= \frac{1}{d-2} \sigma_{(k)}^2\nonumber\\
&+ \left[\frac{1}{(d-2)^2(d-4)}({D}_a \theta_{(k)} - {w}_a  \theta_{(k)} - {R}_{ka})^2+\frac{1}{2(d-2)(d-4)}( D_a \sigma^{(k)}_{bc})^2\right]z^2 + \cdots \label{eq-ssquared},
\end{align}
where we have dropped terms of order $z^2$ which vanish when $\sigma^{(k)}_{ab} =0$. As predicted, these terms are the conformally invariant contributions to $(\delta \bar{X})^2$.

In order to access the state-dependent part of $s^2$ we need the terms in \eqref{eq-ssquared} to vanish. Note that in $d=3$ this always happens. In that case there is no $z^2$ term, and $\sigma^{(k)}_{ab} =0$ always. Though our expression is singular in $d=4$, comparing to (\ref{eq:deltax}) shows that actually the term in brackets above is essentially the same as the $z^2\log z$ term in $\delta \bar{X}$. We already noted that this term was conformally invariant, so this is expected. The difference now is that we no longer need $\theta_{(k)} =0$ in order to get to the QNEC in $d=4$. In $d=5$ the geometric conditions for the state-independent parts of $s^2$ to vanish are identical to those for $d=4$, whereas in the $(\delta \bar{X})^2$ analysis we found that extra conditions were necessary. These were relics of the choice of conformal frame. Finally, for $d>5$ there will be additional state-independent terms that we have not analyzed, but the results we have will still hold.

\paragraph{Conformal QNEC}
Now we analyze the state-dependent part of $s^2$ at order $z^{d-2}$. When all of the state-independent parts vanish, the state-dependent part is given by the conformal transformation of the QNEC. This is easily computed as follows:
\begin{equation}
	L^{-2}\left. s^2 \right|_{z^{d-2}} = \frac{16\pi G_N}{\eta dL^{d-1}}\left[2 \pi \langle \hat{T}_{ij}\rangle k^ik^j - \delta \left( \frac{k^i}{\sqrt{h}} \frac{\delta \hat{S}_{\rm ren}}{\delta X^i(y)} \right) - \frac{d}{2}\theta_{(k)} \left( \frac{k^i}{\sqrt{h}} \frac{\delta \hat{S}_{\rm ren}}{\delta X^i(y)} \right)\right].
\end{equation}
Of course, one would like to replace $\hat{T}_{ij}$ with $T_{ij}$ and $\hat{S}_{\rm ren}$ with $S_{\rm ren}$. When $d$ is odd this is straightforward, as these quantities are conformally invariant. However, when $d$ is even there are anomalies that will contribute, leading to extra geometric terms in the conformal QNEC~\cite{Graham:1999pm,Koeller:2015qmn}.


\section{Connection to Quantum Focusing}


\subsection{The Quantum Foscusing Conjecture \label{Section 3.1}}

We start by reviewing the statement of the QFC \cite{Bousso:2015mna,Leichenauer:2017bmc} before moving on to its connection to EWN and the QNEC. Consider a codimension-two Cauchy-splitting (i.e. entangling) surface $\Sigma$ and a null vector field $k^{i}$ normal to $\Sigma$. Denote by $\mathcal{N}$ the null surface generated by $k^{i}$. The generalized entropy, $S_{\rm gen}$, associated to $\Sigma$ is given by 
\begin{align}
S_{\rm gen} = \langle S_{\rm grav}\rangle+ S_{\rm ren} \label{GenEntropy}
\end{align}
where $S_{\rm grav}$ is a state-independent local integral on $\Sigma$ and $S_{\rm ren}$ is the renormalized von Neumann entropy of the interior (or exterior of $\Sigma$. The terms in $S_{\rm grav}$ are determined by the low-energy effective action of the theory in a well-known way~\cite{Dong:2013qoa}. Even though $\langle S_{\rm grav}\rangle$ and $S_{\rm ren}$ individually depend on the renormalization scheme, that dependence cancels out between them so that $S_{\rm gen}$ is scheme-independent.

The generalized entropy is a functional of the entangling surface $\Sigma$, and the QFC is a statement about what happens when we vary the shape of $\Sigma$ be deforming it within the surface $\mathcal{N}$. Specifically, consider a one-parameter family $\Sigma(\lambda)$ of cuts of $\mathcal{N}$ generated by deforming the original surface using the vector field $k^i$. Here $\lambda$ is the affine parameter along the geodesic generated by $k^{i}$ and $\Sigma(0) \equiv \Sigma$. To be more precise, let $y^a$ denote a set of intrinsic coordinates for $\Sigma$, let $h_{ab}$ be the induced metric on $\Sigma$, and let $X^i(y,\lambda)$ be the embedding functions for $\Sigma(\lambda)$. With this notation, $k^i = \partial_\lambda X^i$. The change in the generalized entropy is given by
\be
\left.\frac{dS_{\rm gen}}{d\lambda}\right|_{\lambda =0} = \int_\Sigma d^{d-2}y~ \frac{\delta S_{\rm gen}}{\delta X^i(y)}\partial_\lambda X^i(y) \equiv \frac{1}{4G_N} \int_\Sigma d^{d-2}y \sqrt{h}\,\Theta[\Sigma,y]
\ee
This defines the quantum expansion $\Theta[\Sigma,y]$ in terms of the functional derivative of the generalized entropy:
\begin{align}
\Theta[\Sigma, y] = 4G_N\frac{k^i(y)}{\sqrt{h}}\frac{\delta S_{\rm gen}}{\delta X^i(y)}.
\end{align}
Note that we have suppressed the dependence of $\Theta$ on $k^i$ in the notation, but the dependence is very simple: if $k^i(y)\to f(y) k^i(y)$, then $\Theta[\Sigma,y] \to f(y)\Theta[\Sigma,y]$.

The QFC is simple to state in terms of $\Theta$. It says that $\Theta$ is non-increasing along the flow generated by $k^i$:  
\begin{align}
0 \geq \frac{d\Theta}{d\lambda} = \int_\Sigma d^{d-2}y~ \frac{\delta \Theta[\Sigma,y]}{\delta X^i(y')} k^i(y'). \label{QFC}
\end{align}
Before moving on, let us make two remarks about the QFC.

First, the functional derivative $\delta \Theta[\Sigma,y]/\delta X^i(y')$ will contain local terms (i.e. terms proportional to $\delta$-functions or derivatives of $\delta$-functions with support at $y=y'$) as well as non-local terms that have support even when $y\neq y'$. $S_{\rm grav}$, being a local integral, will only contribute to the local terms of $\delta \Theta[\Sigma,y]/\delta X^i(y')$. The renormalized entropy $S_{\rm ren}$ will contribute both local and non-local terms. The non-local terms can be shown to be nonpositive using strong subadditivity of the entropy~\cite{Bousso:2015mna}, while the local terms coming from $S_{\rm ren}$ are in general extremely difficult to compute.

Second, and more importantly for us here, the QFC as written in \eqref{QFC} does not quite make sense. We have to remember that $S_{\rm grav}$ is really an operator, and its expectation value $\langle S_{\rm grav} \rangle$ is really the thing that contributes to $\Theta$. In order to be well-defined in the low-energy effective theory of gravity, this expectation value must be smeared over a scale large compared to the cutoff scale of the theory. Thus when we write an inequality like \eqref{QFC}, we are implicitly smearing in $y$ against some profile. The profile we use is arbitrary as long as it is slowly-varying on the cutoff scale. This extra smearing step is necessary to avoid certain violations of \eqref{QFC}, as we will see below~\cite{Leichenauer:2017bmc}.


\subsection{QNEC from QFC}\label{sec:qnecqfc}

In this section we will explicitly evaluate the QFC inequality, \eqref{QFC}, and derive the QNEC in curved space from it as a nongravitational limit. We consider theories with a gravitational action of the form  
\begin{align}
I_{\rm grav} = \frac{1}{16\pi G_N}\int \sqrt{g} \left(R + \ell^2 \lambda_1 R^2 + \ell^2\lambda_2 R_{ij}R^{ij} + \ell^2\lambda_{\rm GB}\mathcal{L}_{\rm GB}\right) \label{action}
\end{align}
where $\mathcal{L}_{GB} = R_{ijmn}^2 -4 R_{ij}^2 + R^2$ is the Gauss-Bonnet Lagrangian. Here $\ell$ is the cutoff length scale of the effective field theory, and the dimensionless couplings $\lambda_1$, $\lambda_2$, and $\lambda_{\rm GB}$ are assumed to be renormalized.

The generalized entropy functional for these theories can be computed using standard replica methods \cite{Dong:2013qoa} and takes the form  
\begin{align}
S_{\rm gen} = \frac{A[\Sigma]}{4G_N} +\frac{\ell^2}{4G_N} \int_\Sigma \sqrt{h}\left[2\lambda_1 R + \lambda_2\left(R_{ij}N^{ij} - \frac{1}{2}K_iK^i\right) + 2\lambda_{\rm GB}r\right] +S_{\rm ren}.\label{eq-SgenExplicit}
\end{align}
Here $A[\Sigma]$ is the area of the entangling surface, $N^{ij}$ is the projector onto the normal space of $\Sigma$, $K^i$ is the trace of the extrinsic curvature of $\Sigma$, and $r$ is the intrinsic Ricci scalar of $\Sigma$.

We can easily compute $\Theta$ by taking a functional derivative of \eqref{eq-SgenExplicit}, taking care to integrate by parts so that the result is proportional to $k^i(y)$ and not derivatives of $k^i(y)$. One finds
\begin{align}
\Theta &= \theta_{(k)} + \ell^2\Biggl[ 2\lambda_1(\theta_{(k)} R + \nabla_k R)\Biggr.
+\lambda_2 \left((D_a -w_a)^2 \theta_{(k)} + K_iK^{iab}K_{ab}^k \right. \\ & + \left.\theta_{(k)} R_{klkl}+\nabla_k R - 2\nabla_l R_{kk} + \theta_{(k)} R_{kl} - \theta_{(l)} R_{kk} + 2K^{kab}R_{ab} \right) \nonumber\\
&- \Biggl. 4\lambda_{GB}\left( r^{ab}K^k_{ab}-\frac{1}{2}r\theta_{(k)}\right)\Biggr] + 4G_N\frac{k^i}{\sqrt{h}}\frac{\delta S_{\rm ren}}{\delta X^i}
\end{align}
Now we must compute the $\lambda$-derivative of $\Theta$. When we do this, the leading term comes from the derivative of $\theta_{(k)}$, which by Raychaudhuri's equation contains the terms $\theta_{(k)}^2$ and $\sigma_{(k)}^2$. Since we are ultimately interested in deriving the QNEC as the non-gravitational limit of the QFC, we need to set $\theta_{(k)} = \sigma^{(k)}_{ab} = 0$ so that the nongravitational limit is not dominated by those terms. So for the rest of this section we will set $\theta_{(k)} = \sigma^{(k)}_{ab} = 0$ at the point of evaluation (but not globally!). Then we find
\begin{align}
\frac{d\Theta}{d\lambda} &= -R_{kk} + 2\lambda_1 \ell^2 \left(\nabla_k^2 R - RR_{kk}\right) \nonumber \\
& + \lambda_2 \ell^2\Big[ 2D_a(w^aR_{kk}) +\nabla_k^2 R -D_aD^aR_{kk} - \frac{d}{d-2}(D_a\theta_{(k)})^2 - 2R_{kb}D^b\theta_{(k)} - 2(D_a\sigma_{bc})^2 \nonumber \\
&  - 2\nabla_k \nabla_l R_{kk} - 2R_{kakb}R^{ab} - \theta_{(l)} \nabla_k R_{kk}\Big]  -2\lambda_{\rm GB}\ell^2\Bigg[\frac{d(d-3)(d-4)}{(d-1)(d-2)^2}RR_{kk} \nonumber\\
& - 4\frac{(d-4)(d-3)}{(d-2)^2}R_{kk}R_{kl}-\frac{2(d-4)}{d-2}C_{klkl}R_{kk} - \frac{2(d-4)}{d-2}R^{ab}C_{akbk}+4C^{kalb}C_{kakb}\Bigg]\nonumber\\
&+ 4G_N\frac{d}{d\lambda}\left(\frac{k^i}{\sqrt{h}}\frac{\delta S_{\rm ren}}{\delta X^i}\right)
\end{align} 
This expression is quite complicated, but it simplifies dramatically if we make use of the equation of motion coming from \eqref{action} plus the action of the matter sector. Then we have $R_{kk} = 8\pi G T_{kk}-H_{kk}$ where \cite{Gurses:2012db}
\begin{align}
\begin{split}
H_{kk} &= 2\lambda_1 \left(RR_{kk} - \nabla_k ^2 R\right) + \lambda_2 \Bigl(2R_{ki k j}R^{i j} - \nabla_k^2 R +2\nabla_k \nabla_l R_{kk} - 2R_{klki}R^i_k \Bigr. \\& \left.+ D_cD^cR_{kk} - 2D_c(w^cR_{kk})-2(D_b\theta_{(k)} + R_{bmkj}P^{mj})R^b_k + \theta_{(l)} \nabla_k R_{kk}\right) \\&
+ 2\lambda_{\rm GB}\left(\frac{d(d-3)(d-4)}{(d-1)(d-2)^2}RR_{kk} - 4\frac{(d-4)(d-3)}{(d-2)^2}R_{kk}R_{kl} - 2\frac{d-4}{d-2}R^{ij}C_{kikj}+C_{kijm}C_k{}^{ijm}\right)
\end{split}
\end{align}
For the Gauss-Bonnet term we have used the standard decomposition of the Riemann tensor in terms of the Weyl and Ricci tensors. Using similar methods to those in Appendix \ref{EWN}, we have also exchanged $k^ik^j \square R_{ij}$ in the $R_{ij}^2$ equation of motion for surface quantities and ambient curvatures.

After using the equation of motion we have the relatively simple formula
\begin{align}\label{eq:S''}
\frac{d\Theta}{d\lambda} &= -\lambda_2 \ell^2\left(\frac{d}{d-2}(D_a\theta_{(k)})^2 + 4R_k^bD_b\theta_{(k)} +2R_{bk}R^b_k+2(D_a\sigma_{bc}^{(k)})^2 \right) \nonumber\\
& +2\lambda_{\rm GB}\ell^2\left(C_{kabc}C_k{}^{abc} - 2C_{kba}{}^bC_{kc}{}^{ac}\right) +4G_N\frac{d}{d\lambda}\left(\frac{k^i}{\sqrt{h}}\frac{\delta S_{\rm ren}}{\delta X^i}\right) - 8\pi G_N \braket{T_{kk}}
\end{align}
The Gauss-Bonnet term agrees with the expression derived in \cite{Fu:2017aa}. However unlike \cite{Fu:2017aa} we have not made any perturbative assumptions about the background curvature.

At first glance it seems like \eqref{eq:S''} does not have definite sign, even in the non-gravitational limit, due to the geometric terms proportional to $\lambda_2$ and $\lambda_{\rm GB}$. The difficulty posed by the Gauss-Bonnet term, in particular, was first pointed out in \cite{Fu:2017ab}. However, this is where we have to remember the smearing prescription mentioned in \S\ref{Section 3.1}. We must integrate \eqref{eq:S''} over a region of size larger than $\ell$ before testing its nonpositivity. The crucial point, used in \cite{Leichenauer:2017bmc}, is that we must also remember to integrate the terms $\theta_{(k)}^2$ and $\sigma_{(k)}^2$ that we dropped earlier over the same region. When we integrate $\theta_{(k)}^2$ over a region of size $\ell$ centered at a point where $\theta_{(k)}=0$, the result is $\xi \ell^2 (D_a\theta_{(k)})^2 + o(\ell^2)$, where $\xi \gtrsim 10$ is a parameter associated with the smearing profile. A similar result holds for $\sigma^{(k)}_{ab}$. Thus we arrive at
\begin{align}\label{eq:S''smeared}
\frac{d\Theta}{d\lambda} &= -\frac{\xi}{d-2} \ell^2 (D_a\theta_{(k)})^2-\xi \ell^2 (D_a\sigma_{bc}^{(k)})^2\nonumber\\
&-\lambda_2 \ell^2\left(\frac{d}{d-2}(D_a\theta_{(k)})^2 + 4R_k^bD_b\theta_{(k)} +2R_{bk}R^b_k+2(D_a\sigma^{(k)}_{bc})^2 \right) \nonumber\\
& +2\lambda_{\rm GB}\ell^2\left(C_{kabc}C_k{}^{abc} - 2C_{kba}{}^bC_{kc}{}^{ac}\right)\nonumber\\
& +4G_N\frac{d}{d\lambda}\left(\frac{k^i}{\sqrt{h}}\frac{\delta S_{\rm ren}}{\delta X^i}\right) - 8\pi G_N \braket{T_{kk}} + o(\ell^2)
\end{align}
Since the size of $\xi$ is determined by the validity of the effective field theory, by construction the terms proportional to $\xi$ in \eqref{eq:S''smeared} dominate over the others. Thus in order to take the non-gravitational limit, we must eliminate these smeared terms.

Clearly we need to be able to choose a surface such that $D_a\theta_{(k)} = D_a\sigma_{bc}^{(k)}=0$. Then smearing $\theta_{(k)}^2$ and $\sigma_{(k)}^2$ would only produce terms of order $\ell^4$ (terms of that order would also show up from smearing the operators proportional to $\lambda_2$ and $\lambda_{\rm GB}$). As explained in \cite{Leichenauer:2017bmc}, this is only possible given certain conditions on the background spacetime at the point of evaluation. We must have
\be
C_{kabc} = \frac{1}{d-2}h_{ab}R_{kc} - \frac{1}{d-2}h_{ac}R_{kb}.
\ee
This can be seen by using the Codazzi equation for $\Sigma$. Imposing this condition, which allows us to set  $D_a\theta_{(k)} = D_a\sigma_{bc}^{(k)}=0$, we then have.
\begin{align}\label{GenEntropyVariation}
\frac{d\Theta}{d\lambda} &=-2\ell^2\left(\lambda_2 + 2\frac{(d-3)(d-4)}{(d-2)^2}\lambda_{\rm GB} \right)R_{bk}R^b_k \nonumber\\
& +4G_N\frac{d}{d\lambda}\left(\frac{k^i}{\sqrt{h}}\frac{\delta S_{\rm ren}}{\delta X^i}\right) - 8\pi G_N \braket{T_{kk}} + o(\ell^3).
\end{align}
This is the quantity which must be negative according to the QFC. In deriving it, we had to assume that $\theta_{(k)} = \sigma^{(k)}= D_a\theta_{(k)} = D_a\sigma^{(k)}_{bc} = 0$.

We make two observations about \eqref{GenEntropyVariation}. First, if we assume that $R_{ka} = 0$ as an additional assumption and take $\ell\to 0$, then we arrive at the QNEC as long as $G_N > o(\ell^3)$. This is the case when $\ell$ scales with the Planck length and $d \leq 5$. These conditions are similar to the ones we found previously from EWN, and below in \S\ref{sec-comparison} we will discuss that in more detail.

The second observation has to do with the lingering possibility of a violation of the QFC due to the terms involving the couplings. In order to have a violation, one would need the linear combination
\be
\lambda_2 + 2\frac{(d-3)(d-4)}{(d-2)^2}\lambda_{\rm GB}
\ee
to be negative. Then if one could find a situation where the first line of \eqref{GenEntropyVariation} dominated over the second, there would be a violation. It would be interesting to interpret this as a bound on the above linear combination of couplings coming from the QFC, but it is difficult to find a situation where the first line of \eqref{GenEntropyVariation} dominates. The only way for $R_{ka}$ to be large compared to the cutoff scale is if $T_{ka}$ is nonzero, in which case we would have $R_{ka} \sim G_N T_{ka}$. Then in order for the first line of \eqref{GenEntropyVariation} to dominate we would need
\be
G_N\ell^2 T_{ka}T^a_k \gg T_{kk}.
\ee
As an example, for a scalar field $\Phi$ this condition would say
\be
G_N\ell^2 (\partial_a\Phi)^2 \gg 1.
\ee
This is not achievable within effective field theory, as it would require the field to have super-Planckian gradients. We leave a detailed and complete discussion of this issue to future work.

\subsection{Scheme-Independence of the QNEC}\label{schemeind}

We take a brief interlude to discuss the issue of the scheme-dependence of the QNEC, which will be important in the following section. It was shown in~\cite{Fu:2017aa}, under some slightly stronger assumptions than the ones we have been using, that the QNEC is scheme-independent under the same conditions where we expect it to hold true. Here we will present our own proof of this fact, which actually follows from the manipulations we performed above involving the QFC.

In this section we will take the point of view of field theory on curved spacetime without dynamical gravity. Then each of the terms in $I_{\rm grav}$, defined above in \eqref{action}, are completely arbitrary, non-dynamical terms we can add to the Lagrangian at will.\footnote{We should really be working at the level of the quantum effective action, or generating functional, for correlation functions of $T_{ij}$~\cite{Fu:2017ab}. The geometrical part has the same form as the classical action $I_{\rm grav}$ and so does not alter this discussion.} Dialing the values of those various couplings corresponds to a choice of {\em scheme}, as even though those couplings are non-dynamical they will still contribute to the definitions of quantities like the renormalized energy-momentum tensor and the renormalized entropy (as defined through the replica trick). The QNEC is scheme-independent if it is insensitive to the values of these couplings.

To show the scheme-independence of the QNEC, we will begin with the statement that $S_{\rm gen}$ is scheme-independent. We remarked on this above, when our context was a theory with dynamical gravity. But the scheme-independence of $S_{\rm gen}$ does not require use of the equations of motion, so it is valid even in a non-gravitational theory on a fixed background. In fact, only once in the above discussion did we make use of the gravitational equations of motion, and that was in deriving \eqref{eq:S''}. Following the same steps up to that point, but without imposing the gravitational equations of motion, we find instead
\begin{align}
\frac{d\Theta}{d\lambda} &= -\lambda_2 \ell^2\left(\frac{d}{d-2}(D_a\theta_{(k)})^2 + 4R_k^bD_b\theta_{(k)} +2R_{bk}R^b_k+2(D_a\sigma_{bc})^2 \right) \nonumber\\
& +2\lambda_{\rm GB}\ell^2\left(C_{kabc}C_k{}^{abc} - 2C_{kba}{}^bC_{kc}{}^{ac}\right) +4G_N\frac{d}{d\lambda}\left(\frac{k^i}{\sqrt{h}}\frac{\delta S_{\rm ren}}{\delta X^i}\right) - k_ik_j \frac{16\pi G_N}{\sqrt{g}} \frac{\delta I_{\rm grav}}{\delta g_{ij}}.
\end{align}
Since the theory is not gravitational, we would not claim that this quantity has a sign. However, it is still scheme-independent.

To proceed, we will impose all of the additional conditions that are necessary to prove the QNEC. That is, we impose $D_b\theta_{(k)} = R_k^b = D_a\sigma_{bc} = 0$, as well as $\theta_{(k)} = \sigma^{(k)}_{ab} =0$, which in turn requires $C_{kabc}=0$. Under these conditions, we learn that the combination
\begin{align}
\frac{d}{d\lambda}\left(\frac{k^i}{\sqrt{h}}\frac{\delta S_{\rm ren}}{\delta X^i}\right) - k_ik_j \frac{4\pi}{\sqrt{g}} \frac{\delta I_{\rm grav}}{\delta g_{ij}}
\end{align}
is scheme-independent. The second term here is one of the contributions to the renormalized $2\pi \langle T_{kk} \rangle$ in the non-gravitational setup, the other contribution being $k_ik_j \frac{4\pi}{\sqrt{g}} \frac{\delta I_{\rm matter}}{\delta g_{ij}}$. But $I_{\rm matter}$ is already scheme-independent in the sense we are discussing, in that it is independent of the parameters appearing in $I_{\rm grav}$. So adding that to the terms we have above, we learn that
\be
\frac{d}{d\lambda}\left(\frac{k^i}{\sqrt{h}}\frac{\delta S_{\rm ren}}{\delta X^i}\right) - 2\pi \langle T_{kk} \rangle
\ee
is scheme-independent. This is what we wanted to show.


\subsection{QFC vs EWN \label{sec-comparison}}

As we have discussed above, by taking the non-gravitational limit of \eqref{GenEntropyVariation} under the assumptions $D_b\theta_{(k)} = R_k^b = D_a\sigma_{bc} = \theta_{(k)} = \sigma^{(k)}_{ab} =0$ we find the QNEC as a consequence of the QFC (at least for $d\leq 5$). And under the same set of geometric assumptions, we found the QNEC as a consequence of EWN in \eqref{eq-EWNQNEC}. The discussion of the previous section demonstrates that these assumptions also guarantee that the QNEC is scheme-independent. So even though these two QNEC inequalities were derived in different ways, we know that at the end of the day they are the same QNEC. It is natural to ask if there is a further relationship between EWN and the QFC, beyond the fact that they give the same QNEC. We will begin to investigate that question in this section.

The natural thing to ask about is the state-independent terms in the QFC and in $(\delta \bar{X})^2$. We begin by writing down all of the terms of $(\delta \bar{X})^2$ in odd dimensions that we have computed: 
\begin{align}
(d-2)L^{-2}(\delta \bar{X}^i)^2 &= \frac{1}{(d-2)}\theta_{(k)}^2 +\sigma_{(k)}^2\nonumber\\
&+z^2\frac{1}{4(d-2)}(D_a\theta_{(k)} +2R_{ka})^2\nonumber\\
&+z^2\frac{1}{(d-2)(d-4)}(D_a\theta_{(k)} + R_{ka})^2+z^2\frac{1}{2(d-4)}(D_a\sigma^{(k)}_{bc})^2\nonumber\\
&+z^2\frac{\kappa}{d-4}\left(C_{kabc}C_{k}^{~abc}-2C_{k~ca}^{~c}C_{k~b}^{~b~a}\right)\nonumber\\
&+\cdots+z^{d-2}\frac{16\pi (d-2)G_{N}}{\eta d L^{d-1}}\left[\langle T_{kk}\rangle- \delta\left(\frac{k^i}{2\pi\sqrt{h}}\frac{\delta S_{\rm ren}}{\delta X^i}\right)\right].
\end{align}
The first line looks like $-\dot\theta$, which would be the leading term in $d\Theta/d\lambda$, except it is missing an $R_{kk}$. Of course, we eventually got rid of the $R_{kk}$ in the QFC by using the equations of motion. Suppose we set $\theta_{(k)} = 0$ and $\sigma_{ab}^{(k)} = 0$ to eliminate those terms, as we did with the QFC. Then we can write $(\delta \bar{X})^2$ suggestively as
\begin{align}
(d-2)L^{-2}(\delta \bar{X}^i)^2 &=z^2\tilde\lambda_2\Big(\frac{d}{(d-2)}(D_a\theta_k)^2 +4R_{k}^aD_a\theta + \frac{4(d-3)}{(d-2)}R_{ka}R_k^a+2(D_a\sigma^{(k)}_{bc})^2\Big)\nonumber\\
&-2z^2\tilde \lambda_{\rm GB}\left(C_{kabc}C_{k}^{~abc}-2C_{k~ca}^{~c}C_{k~b}^{~b~a}\right)\nonumber\\
&+\cdots+8\pi \tilde G_{N}\langle T_{kk}\rangle- 4\tilde G_N\delta\left(\frac{k^i}{\sqrt{h}}\frac{\delta S_{\rm ren}}{\delta X^i}\right).
\end{align}
where
\begin{align}
\tilde G_N &= G_N \frac{2(d-2) z^{d-2}}{\eta dL^{d-1}},\\
\tilde \lambda_2 &= \frac{1}{4(d-4)},\\
\tilde \lambda_{\rm GB} &= -\frac{\kappa}{2(d-4)}.
\end{align}
Written this way, it almost seems like $(d-2)L^{-2}(\delta \bar{X}^i)^2 \sim -d\Theta/d\lambda$ in some kind of model gravitational theory. One discrepancy is in the coefficient of the $R_{ka}R^{ka}$ term, unless $d=4$. It is also intriguing that the effective coefficients $\tilde G_N$,  $\tilde \lambda_2$, and $\tilde \lambda_{\rm GB}$ are close to, but not exactly the same as, the effective braneworld induced gravity coefficients found in~\cite{Myers:2013lva}. This is clearly something that deserves further study.


\section{Discussion}\label{discussion}


We have displayed a strong similarity between the state-independent inequalities in the QFC and the state-independent inequalities from EWN. We now discuss several possible future directions and open questions that follow naturally from these results.

\subsection{Bulk Entropy Contributions}\label{sec-bulkent}

We ignored the bulk entropy $S_{\rm bulk}$ in this work, but we know that it produces a contribution to CFT entropy~\cite{Faulkner:2013ana} and plays a role in the position of the extremal surface~\cite{Engelhardt:2014gca,Dong:2017aa}. The bulk entropy contributions to the entropy are subleading in $N^2$ and do not interfere with the gravitational terms in the entropy. We could include the bulk entropy as a source term in the equations determining $\bar{X}$, which could lead to extra contributions to the $X_{(n)}$ coefficients. However, it does not seem possible for the bulk entropy to have an effect on the state-independent parts of the extremal surface, namely on  $X_{(n)}$ for $n<d$, which means the bulk entropy would not affect the conditions we derived for when the QNEC should hold.

Another logical possibility is that the bulk entropy term could affect the statement of the QNEC itself, meaning that the schematic form $T_{kk} -S''$ would be altered. This would be problematic, especially given that the QFC always produces a QNEC of that same form. It was argued in \cite{Akers:2016aa} that this does not happen, and that argument holds here as well.

\subsection{Smearing of EWN}

We were careful to include a smearing prescription for defining the QFC, and it was an important ingredient in the analysis of \S\ref{sec:qnecqfc}. But what about smearing of EWN? Of course, the answer is that we {\em should} smear EWN appropriately, but as we will see now it would not make a difference to our analysis,

The issue is that the bulk theory is a low-energy effective theory of gravity with a cutoff scale $\ell$, and the quantities that we use to probe EWN, like $(\delta \bar{X})^2$, are operators in that theory. As such, these operators need to be smeared over a region of proper size $\ell$ on the extremal surface. Of course, due to the warp factor, such a region has coordinate size $z\ell/L$. We can ask what effect such a smearing would have on the inequality $(\delta \bar{X})^2$.

When we performed our QNEC derivation, we assumed that $\theta_{(k)}=0$ at the point of evaluation, so that the $\theta_{(k)}^2$ term in $\left.(\delta \bar{X})^2\right|_{z^0}$ would not contribute. However, after smearing this term would contribute a term of the form $\ell^2(D_a\theta_{(k)})^2/L^2$ to $\left.(\delta \bar{X})^2\right|_{z^2}$. But we already had such a term at this order, so all this does is shift the coefficient. Furthermore, the coefficient is shifted only by an amount of order $\ell^2/L^2$. If the cutoff $\ell$ is of order the Planck scale, then this is suppressed in powers of $N^2$. In other words, this effect is negligible for the analysis. A similar statement applies for $\sigma^{(k)}_{ab}$. So in summary, EWN should be smeared, but the analysis we performed was insensitive to it.

\subsection{Future Work}

There are a number of topics that merit investigation in future work. We will touch on a few of them to finish our discussion.

\paragraph{Relevant Deformations}

Perhaps the first natural extension of our work is to include relevant deformations in the EWN calculation. There are a few reasons why this is interesting. First, one would like to test the continued correspondence between the QFC and EWN when it comes to the QNEC. The QFC arguments do not care whether relevant deformations are turned on, so one would expect that the same is true in EWN. This is indeed the case when the boundary theory is formulated on flat space~\cite{Koeller:2015qmn}, and one would expect similar results to hold when the boundary is curved.

Another reason to add in relevant deformations is to test the status of the Conformal QNEC when the theory is not a CFT. To be more precise, the $(\delta \bar{X})^2$ and $s^2$ calculations we performed differed by a Weyl transformation on the boundary, and since our boundary theory was a CFT this was a natural thing to do. When the boundary theory is not a CFT, what is the relationship between $(\delta \bar{X})^2$ and $s^2$? One possibility, perhaps the most likely one, is that they simply reduce to the same inequality, and the Conformal QNEC no longer holds. It would be good to know the answer.

Finally, and more speculatively, having a relevant deformation turned on when the background is curved allows for interesting state-independent inequalities from EWN. We saw that for a CFT the state-independent terms in both $(\delta \bar X)^2$ and $s^2$ were trivially positive. Perhaps when a relevant deformation is turned on then more nontrivial things might happen, such as the possibility of a $c$-theorem hiding inside of EWN. We are encouraged by the similarity of inequalities used in recent proofs of the $c$-theorems to inequalities obtained from EWN~\cite{Casini:2017vbe}.

\paragraph{Higher Dimensions}

Another pressing issue is extending our results to $d=6$ and beyond. This is an algebraically daunting task using the methods we have used for $d\leq 5$. Considering the ultimate simplicity of our final expressions, especially compared to the intermediate steps in the calculations, it is likely that there are better ways of formulating and performing the analyses we performed here. It is hard to imagine performing the full $d=6$ analysis without such a simplification.

\paragraph{Further Connections Between EWN and QFC}

Despite the issues outlined in \S\ref{sec-comparison}, we are still intrigued by the similarities between EWN and the QFC. It is extremely natural to couple the boundary theory in AdS/CFT to gravity using a braneworld setup~\cite{Randall:1999vf,Verlinde:1999fy,Gubser:1999vj,Myers:2013lva}. Upon doing this, one can formulate the QFC on the braneworld. However, at the same time near-boundary EWN becomes lost, or at least changes form: extremal surfaces anchored to a brane will in general not be orthogonal to the brane, and in that case a null deformation on the brane will induce a timelike deformation of the extremal surface in the vicinity of the brane. Of course, one has to be careful to take into account the uncertainty in the position of the brane, which complicates things. We hope that such an analysis could serve to unify the QFC with EWN, or at least illustrate their relationship with each other.

\paragraph{Conformal QNEC from QFC}

While we emphasized the apparent similarity between the EWN-derived inequality $(\delta \bar{X})^2 \geq 0$ and the QFC, the stronger EWN inequality $s^2 \geq 0$ is nowhere to be found in the QFC discussion. It would be inteesting to see if there was some direct QFC-like way to derive the Conformal QNEC (rather than first deriving the ordinary QNEC and then performing a Weyl transformation). In particular, the Conformal QNEC applies even in cases where $\theta_{(k)}$ is nonzero, while in those cases the QFC is dominated by classical effects. Perhaps there is a useful change of variables that one can do in the semiclassical gravity when the matter sector is a CFT which makes the Conformal QNEC manifest from the QFC point of view. This is worth exploring.

\section*{Acknowledgements}
It is a pleasure to thank R.~Bousso, J.~Koeller, and A.~Wall for discussions. Our work is supported in part by the Berkeley Center for Theoretical Physics, by the National Science Foundation (award numbers 1521446, and 1316783), by FQXi, and by the US Department of Energy under contract DE-AC02-05CH11231.


\appendix

\section{Notation and Definitions}\label{notation}

\subsection{Basic Notation}

\paragraph{Notation for basic bulk and boundary quantities}
\begin{itemize}
\item Bulk indices are $\mu, \nu,\ldots$.
\item Boundary indices are $i,j,\ldots$. Then $\mu = (z,i)$.
\item We assume a Fefferman--Graham form for the metric: $ds^2 = \frac{L^2}{z^2}(dz^2 + \bar{g}_{ij}dx^idx^j)$.
\item The expansion for $\bar{g}_{ij}(x,z)$ at fixed $x$ is 
\be
\bar{g}_{ij} = g_{ij}^{(0)} + z^2g_{ij}^{(2)} + z^4 g_{ij}^{(4)} +\cdots +z^d\log z g_{ij}^{(d,{\rm log})}+ z^dg_{ij}^{(d)} + \cdots.
\ee
The coefficients $g_{ij}^{(n)}$ for $n<d$ and $g_{ij}^{(d,{\rm log})}$ are determined in terms of $g_{ij}^{(0)}$, while $g_{ij}^{(d)}$ is state-dependent and contains the energy-momentum tensor of the CFT. If $d$ is even, then $g_{ij}^{(d,{\rm log})}=0$. To avoid clutter we will often write $g_{ij}^{(0)}$ simply as $g_{ij}$. Unless otherwise indicated, $i,j$ indices are raised and lowered by $g_{ij}^{(0)}$.
\item We use $\mathcal{R}$, $\mathcal{R}_{\mu\nu}$, $\mathcal{R}_{\mu\nu\rho\sigma}$ to denote bulk curvature tensors, and $R$, $R_{ij}$, $R_{ijmn}$ to denote boundary curvature tensors.
\end{itemize}
\paragraph{Notation for extremal surface and entangling surface quantities}
\begin{itemize}
\item Extremal surface indices are $\alpha, \beta,\ldots$.
\item Boundary indices are $a,b,\ldots$. Then $\alpha = (z,a)$.
\item The extremal surface is parameterized by functions $\bar{X}^\mu(z,y^a)$. We choose a gauge such that $X^z=z$, and expand the remaining coordinates as 
\be \bar{X}^i = X^i_{(0)} + z^2X^i_{(2)} + z^4X^i_{(4)} + \cdots +z^d\log z X^i_{(d,{\rm log})}+z^dX^i_{(d)} + \cdots.
\ee 
The coefficients $X^i_{(n)}$ for $n<d$ and $X^i_{(d,{\rm log})}$ are determined in terms of $X^i_{(0)}$ and $g^{(0)}_{ij}$, while $X^i_{(d)}$ is state-dependent and is related to the renormalized entropy of the CFT region.
\item The extremal surface induced metric will be denoted $\bar{h}_{\alpha \beta}$ and gauge-fixed so that $\bar{h}_{za}=0$.
\item The entangling surface induced metric will be denoted $h_{ab}$.
\item Note that we will often want to expand bulk quantities in $z$ at fixed $y$ instead of fixed $x$. For instance, the bulk metric at fixed $y$ is
\begin{align}
\bar{g}_{ij}(y,z) &= \bar{g}_{ij}(\bar{X}(z,y),z) = \bar{g}_{ij}(X_{(0)}(y)+z^2X_{(2)}(y)+\cdots,z)\nonumber\\
&=g_{ij}^{(0)} + z^2\left(g^{(2)}_{ij}+ X_{(2)}^m\partial_mg_{ij}^{(0)}\right)+\cdots
\end{align}
Similar remarks apply for things like Christoffel symbols. The prescription is to always compute the given quantity as a function of $x$ first, the plug in $\bar{X}(y,z)$ and expand in a Taylor series.
\end{itemize}

\subsection{Intrinsic and Extrinsic Geometry}

Now will introduce several geometric quantities, and their notations, which we will need. First, we define a basis of surface tangent vectors by
\be
e_a^i = \partial_a X^i.
\ee
We will also make use of the convention that ambient tensors which are not inherently defined on the surface but are written with surface indices ($a$, $b$, etc.) are defined by contracting with $e_a^i$. For instance:
\be
g^{(2)}_{aj} = e_a^ig^{(2)}_{ij}.
\ee
We can form the surface projector by contracting the surface indices on two copies of $e_a^i$:
\be
P^{ij} = h^{ab} e^i_a e^j_b = e^i_a e^{ja}.
\ee
We introduces a surface covariant derivative $D_a$ that acts as the covariant derivative on both surface and ambient indices. So it is compatible with both metrics:
\be
D_ah_{bc} = 0 = D_a g_{ij}.
\ee
Note also that when acting on objects with only ambient indices, we have the relationship
\be
D_a V^{ij\cdots}_{pq\cdots} = e_a^m\nabla_mV^{ij\cdots}_{pq\cdots},
\ee
where $\nabla_i$ is the ambient covariant derivative compatible with $g_{ij}$.

The extrinsic curvature is computed by taking the $D_a$ derivative of a surface basis vector:
\be
K^i_{ab} = -D_ae_b^i = -\partial_ae_b^i + \gamma_{ab}^ce_b^i - \Gamma^i_{ab}.
\ee
Note the overall sign we have chosen. Here $\gamma_{ab}^c$ is the Christoffel symbol of the metric $h_{ab}$, and the lower indices on the $\Gamma$ symbol were contracted with two basis tangent vectors to turn them into surface indices. Note that $K_{ab}^i$ is symmetric in its lower indices. It is an exercise to check that it is normal to the surface in its upper index:
\be
e_{ic}K^i_{ab} = 0.
\ee
The trace of the extrinsic curvature is denoted by $K^i$:
\be
K^i = h^{ab}K^i_{ab}.
\ee
Below we will introduce the null basis of normal vectors $k^i$ and $l^i$. Then we can define expansion $\theta_{(k)}$ ($\theta_{(l)}$) and shear $\sigma^{(k)}_{ab}$ ($\sigma^{(l)}_{ab}$) as the trace and traceless parts of $k_iK^i_{ab}$ ($l_iK^i_{ab}$), respectively.

There are a couple of important formulas involving the extrinsic curvature. First is the Codazzi Equation, which can be computed from the commutator of covariant derivatives:
\begin{align}
\begin{split}
D_cK_{ab}^i - D_bK_{ac}^i &= (D_bD_c-D_cD_b)e_a^i\\
&=R^i_{~abc} - r^d_{~abc}e^i_d.
\end{split}
\end{align}
Here $R^i_{~abc}$ is the ambient curvature (appropriately contracted with surface basis vectors), while $r^d_{~abc}$ is the surface curvature. We can take traces of this equation to get others. Another useful thing to do is contract this equation with $e_d^i$ and differentiate by parts, which yields the Gauss--Codazzi equation:
\begin{align}
K_{cdi}  K_{ab}^i - K_{bdi}K_{ac}^i &= R_{dabc} - r_{dabc}.
\end{align}
Various traces of this equation are also useful.

\subsection{Null Normals $k$ and $l$}\label{sec-kl}

A primary object in our analysis is the bull vector $k^i$, which is orthogonal to the entangling surface and gives the direction of the surface deformation. It will be convenient to also introduce the null normal $l^i$, which is defined so that $l_i k^i = +1$. This choice of sign is different from the one that is usually made in these sorts of analysis, but it is necessary to avoid a proliferation of minus signs. With this convention, the projector onto the normal space of the surface is
\be
N^{ij} \equiv g^{ij} - P^{ij} = k^il^j + k^jl^i = 2k^{(i}l^{j)}.
\ee

As we did with the tangent vectors $e_a^i$, we will introduce a shorthand notation to denote contraction with $k^i$ or $l^i$: any tensor with $k$ or $l$ index means it has been contracted with $k^i$ or $l^i$. As such we will avoid using the letters $k$ and $l$ as dummy indices. For instance.
\be
R_{kl} \equiv k^i l^j R_{ij}.
\ee

Another quantity associated with $k^i$ and $l^i$ is the normal connection $w^a$, defined through
\be
w_a \equiv l_i D_a k^i. 
\ee
With this definition, the tangent derivative of $k^i$ can be shown to be
\be
D_a k^i = w_a k^i + K^k_{ab}e^{bi},
\ee
which is a formula that is used repeatedly in our analysis.

At certain intermediate stages of our calculations it will be convenient to define extensions of $k^i$ and $l^i$ off of the entangling surface, so here we will define such an extension. Surface deformations in both the QNEC and QFC follow geodesics generated by $k^i$, so it makes sense to define $k^i$ to satisfy the geodesic equation:
\be
\nabla_k k^i = 0.
\ee
However, we will {\em not} define $l^i$ by parallel transport along $k^i$. It is conceptually cleaner to maintain the orthogonality of $l^i$ to the surface even as the surface is deformed along the geodesics generated by $k^i$. This means that $l^i$ satisfies the equation
\be
\nabla_k l^i = -w^ae^i_a.
\ee
These equations are enough to specify $l^i$ and $k^i$ on the null surface formed by the geodesics generated by $k^i$. To extend $k^i$ and $l^i$ off of this surface, we specify that they are both parallel-transported along $l^i$. In other words, the null surface generated by $k^i$ forms the initial condition surface for the vector fields $k^i$ and $l^i$ which satisfy the differential equations
\be
\nabla_l k^i = 0,~~~~\nabla_l l^i = 0~.
\ee
This suffices to specify $k^i$ an $l^i$ completely in a neighborhood of the original entangling surface. Now that we have done that, we record the commutator of the two fields for future use:
\be
[k,l]^i =\nabla_k l^i - \nabla_l k^i = -w^ce_c^i.
\ee

\section{Surface Variations}\label{Var}

Most of the technical parts of our analysis have to do with variations of surface quantities under the deformation $X^i \to X^i + \delta X^i$ of the surface embedding coordinates. Here $\delta X^i$ should be interpreted a vector field defined on the surface. In principle it can include both normal and tangential components, but since tangential components do not actually correspond to physical deformations of the surface we will assume that $\delta X^i$ is normal. The operator $\delta$ denotes the change in a quantity under the variation. In the case where $\delta X^i = \partial_\lambda X^i$, which is the case we are primarily interested in, $\delta$ can be identified with $\partial_\lambda$. With this in mind, we will always impose the geodesic equation on $k^i$ whenever convenient. In terms of the notation we are introducing here, this is
\be
\delta k^i = - \Gamma^i_{kk}.
\ee

To make contact with the main text, we will use the notation $k^i \equiv \delta X^i$, and assume that $k^i$ is null since that is ultimately the case we care about. Some of the formulas we discuss below will not depend on the fact that $k^i$ is null, but we will not make an attempt to distinguish them.

\paragraph{Ambient Quantities}

For ambient quantities, like curvature tensors, the variation $\delta$ can be interpreted straightforwardly as $k^i\partial_i$ with no other qualification. Thus we can freely use, for instance, the ambient covariant derivative $\nabla_k$ to simplify the calculations of these quantities. Note that $\delta$ itself is {\rm not} the covariant derivative. As defined, $\delta$ is a coordinate dependent operator. This may be less-than-optimal from a geometric point of view, but it has the most conceptually straightforward interpretation in terms of the calculus of variations. In all of the variational formulas below, then, we will see explicit Christoffel symbols appear. Of course, ultimately these non-covariant terms must cancel out of physical quantities. That they do serves as a nice check on our algebra.

\paragraph{Tangent Vectors}

The most fundamental formula is that of the variation of the tangent vectors $e_a^i \equiv \partial_a X^i$. Directly from the definition, we have
\be
\delta e_a^i = \partial_a k^i = D_a k^i - \Gamma^i_{ak} = w_a k^i + K^k_{ab}e^{bi} - \Gamma^i_{ak}.
\ee
This formula, together with the discussion of how ambient quantities transform, can be used together to compute the variations of many other quantities.

\paragraph{Intrinsic Geometry and Normal Vectors}

The intrinsic metric variation is easily computed from the above formula as
\be
\delta h_{ab} = 2K^k_{ab}.
\ee
From here we can find the variation of the tangent projector, for instance:
\begin{align}
\delta P^{ij} &= \delta h^{ab} e_a^i e_b^j + 2h^{ab}e_a^{(i} \partial_b k^{j)}\nonumber \\
&= -2K_k^{ab}e_a^ie_b^j + 2h^{ab}e_a^{(i} D_b k^{j)}- 2h^{ab}e_a^{(i} \Gamma_{bk}^{j)}\nonumber \\
&= 2w^a e_a^{(i} k^{j)}- 2h^{ab}e_a^{(i} \Gamma_{bk}^{j)}.
\end{align}
Notice that the second line features a derivative of $k^i = \delta X^i$. In a context where we are taking functional derivatives, such as when computing equations of motion, this term would require integration by parts. We can write the last line covariantly as
\begin{align}
\nabla_k P^{ij} &= 2w^a e_a^{(i} k^{j)}.
\end{align}

Earlier we saw that $l^i$ satisfied the equation $\nabla_k l^i = -w^ae_a^i$ as a result of keeping $l^i$ orthogonal to the surface even as the surface is deformed. In the language of this section, this is seen by the following manipulation:
\be
e^i_a\delta l_i = -l_i \partial_a k^i = - w_a  - \Gamma_{ak}^l.
\ee
Again, note the derivative of $k^i$. It is easy to confirm that represents the only nonzero component of $\nabla_kl^i$.

The normal connection $w_a = l^i D_a k_i$ makes frequent appearances in our calculations, and we will need to know its variation. We can calculate that as follows:
\begin{align}
\delta w_a &= \delta l^i D_a k_i +l^i \partial_a \delta k_i - l^i \delta \Gamma_{ji}^n  e_a^j k_n- l^i \Gamma_{ji}^n \partial_ak^j k_n-l^i \Gamma_{ji}^ne_a^j \delta k_n \nonumber\\
&=\nabla_k l^i D_a k_i +R_{klak} \nonumber\\
&=-w^cK_{ac} +R_{klak}.\label{eq-wvar}
\end{align}

\paragraph{Extrinsic Curvatures}

The simplest extrinsic curvature variation is that of the trace of the extrinsic curvature
\begin{align}
\delta K^i &= -K^m\Gamma_{mk}^i-D_a D^a k^i - R^i_{mkj}P^{mj}   + \left(2D^a (K_{ad}^k)- D_d (K^k) \right) e^{di} - 2 K^{ab}_k K^i_{ab}
\end{align}
Note that the combination $\delta K^i +K^k\Gamma_{km}^ik^m$ is covariant, so it makes sense to write
\begin{align}
\nabla_k K^i &= -D_a D^a k^i - R^i_{mkj}P^{mj}   + \left(2D^a (K_{ad}^k)- D_d (K^k) \right) e^{di} - 2 K^{ab}_k K^i_{ab}
\end{align}
This formula is noteworthy because of the first term, which features derivatives of $k^i = \delta X^i$. This is important because when $K^i$ occurs inside of an integral and we want to compute the functional derivative then we have to first integrate by parts to move those derivatives off of $k^i$. This issue arises when computing $\Theta$ as in the QFC, for instance.

We can contract the previous formulas with $l^i$ and $k^i$ to produce other useful formulas. For instance, contracting with $k^i$ leads to 
\be
\delta K^k = - K^{kab} K^k_{ab}- R_{kk},
\ee
which is nothing but the Raychaudhuri equation.

The variation of the full extrinsic curvature $K^i_{ab}$ is quite complicated, but we will not needed. However, its contraction with $k^i$ will be useful and so we record it here:
\be
k_i\delta K^i_{ab} = -K^j_{ab}\Gamma^m_{jn}k_mk^n - k_iD_aD_bk^i - R_{kakb}.
\ee


\section{$z$-Expansions}\label{zexp}

\subsection{Bulk Metric}
We are focusing on bulk theories with gravitational Lagrangians
\be
\mathcal{L} = \frac{1}{16\pi G_N} \left(\frac{d(d-1)}{\tilde L^2} + \mathcal{R} + \ell^2 \lambda_1 \mathcal{R}^2 + \ell^2 \lambda_2 \mathcal{R}_{\mu\nu}^2+ \ell^2 \lambda_{\rm GB}\mathcal{L}_{\rm GB} \right).
\ee
where $\mathcal{L}_{GB} = \mathcal{R}_{\mu\nu\rho\sigma}^2 -4 \mathcal{R}_{\mu\nu}^2 + \mathcal{R}^2$ is the Gauss-Bonnet Lagrangian, $\ell$ is the cutoff length scale of the bulk effective field theory, and the couplings $\lambda_1$, $\lambda_2$, and $\lambda_{\rm GB}$ are defined to be dimensionless. We have decided to include $\mathcal{L}_{GB}$ as part of our basis of interactions rather than $\mathcal{R}_{\mu\nu\rho\sigma}^2$ because of certain nice properties that the Gauss-Bonnet term has, but this is not important.

We recall that the Fefferman--Graham form of the metric is defined by
\be
ds^2 = \frac{1}{z^2}(dz^2 +\bar{g}_{ij}dx^idx^j),
\ee
where $\bar{g}_{ij}(x,z)$ is expanded as a series in $z$:
\be
\bar{g}_{ij} = g_{ij}^{(0)} + z^2g_{ij}^{(2)} + z^4 g_{ij}^{(4)} +\cdots +z^d\log z g_{ij}^{(d,{\rm log})}+ z^dg_{ij}^{(d)} + \cdots.
\ee
In principle, one would evaluate the equation of motion from the above Lagrangian using the Fefferman--Graham metric form as an ansatz to compute these coefficients. The results of this calculation are largely in the literature, and we quote them here. To save notational clutter, in this section we will set $g_{ij} = g_{ij}^{(0)}$.

The first nontrivial term in the metric expansion is independent of the higher-derivative couplings, and in fact is completely determined by symmetry~\cite{Imbimbo:1999bj}:
\be
g^{(2)}_{ij} = -\frac{1}{d-2}\left(R_{ij} - \frac{1}{2(d-1)}Rg_{ij}\right).
\ee
The next term is also largely determined by symmetry, except for a pair of coefficients~\cite{Imbimbo:1999bj}. We are only interested in the $kk$-component of $g^{(4)}_{ij}$, and where one of the coefficients drops out. The result is
\begin{align}
g^{(4)}_{kk} &= \frac{1}{d-4}\left[\kappa C_{kijm}C_k^{~ijm}+\frac{1}{ 8(d-1)}\nabla_k^2R - \frac{1}{4(d-2)}k^ik^j  \square R_{ij} \right.\nonumber\\
&\left.-\frac{1}{2(d-2)}R^{ij}R_{kikj}+\frac{d-4}{2(d-2)^2}R_{ki}R_k^i + \frac{1}{(d-1)(d-2)^2}RR_{kk} \right],\label{eq-g4kk}
\end{align}
where $C_{ijmn}$ is the Weyl tensor and
\be\label{eq-kappa}
\kappa = -\lambda_{GB}\frac{\ell^2}{L^2}\left(1+O\left(\frac{\ell^2}{L^2}\right)\right).
\ee
In $d=4$ we will need an expression for $g^{(4,{\rm log})}_{kk}$ as well. One can check that this is obtainable from $g^{(4)}_{kk}$ by first multiplying by $4-d$ and then setting $d\to 4$. We record the answer for future reference:
\begin{align}\label{eq-glog}
g^{(4, {\rm log})}_{kk} &=-\left[\kappa C_{kijm}C_k^{~ijm}+\frac{1}{ 24}\nabla_k^2R - \frac{1}{8}k^ik^j  \square R_{ij} -\frac{1}{4}R^{ij}R_{kikj} + \frac{1}{12}RR_{kk} \right].
\end{align}

\subsection{Extremal Surface Coordinates}\label{ExSurf}

The extremal surface position is determined by extremizing the generalized entropy functional~\cite{Engelhardt:2014gca, Dong:2017aa}:
\begin{align}
S_{\rm gen} = \frac{1}{4G_N} \int \sqrt{\bar{h}}\left[1 + 2\lambda_1\ell^2 \mathcal{R} + \lambda_2\ell^2\left(\mathcal{R}_{\mu\nu}\mathcal{N}^{\mu\nu} - \frac{1}{2}\mathcal{K}_\mu\mathcal{K}^\mu\right) + 2\lambda_{\rm GB}\ell^2\bar{r}\right] +S_{\rm bulk}.
\end{align}
Here we are using $\mathcal{K}^i$ to denote the extrinsic curvature and $\bar{r}$ the intrinsic Ricci scalar of the surface.

The equation of motion comes from varying $S_{\rm gen}$ and is (ignoring the $S_{\rm bulk}$ term for simplicity)
\begin{align}
0&=\mathcal{K}^\mu\left[1 + 2\lambda_1\ell^2 \mathcal{R} + \lambda_2\ell^2\left(\mathcal{R}_{\rho\nu}\mathcal{N}^{\rho\nu} - \frac{1}{2}\mathcal{K}_\rho\mathcal{K}^\rho\right) + 2\lambda_{\rm GB}\ell^2\bar{r}\right]+2\lambda_1\ell^2 \nabla^\mu\mathcal{R}  \nonumber\\
&+ \lambda_2\ell^2\Big(\mathcal{N}^{\rho\nu}\nabla^\mu \mathcal{R}_{\rho\nu}+2\mathcal{P}^{\rho\nu}\nabla_\rho\mathcal{R}_{\nu}^\mu-2\mathcal{R}_\rho^\mu\mathcal{K}^\rho + 2\mathcal{K}^{\mu\alpha\beta}\mathcal{R}_{\alpha\beta}+D_\alpha D^\alpha\mathcal{K}^\mu\nonumber \\
 &+ \mathcal{K}^\rho\mathcal{R}_{\mu\sigma\rho\nu}\mathcal{P}^{\nu\sigma}+2\mathcal{K}^{\mu\alpha\beta}\mathcal{K}_\nu\mathcal{K}^\nu_{\alpha\beta}\Big)- 4\lambda_{\rm GB}\ell^2 \bar{r}^{\alpha\beta}\mathcal{K}_{\alpha\beta}^\mu.
\end{align}
This equation is very complicated, but since we are working in $d\leq 5$ dimensions we only need to solve perturbatively in $z$ for $X_{(2)}^i$ and $X_{(4)}^i$\footnote{It goes without saying that these formulas are only valid for $d>2$ and $d>4$, respectively.}. Furthermore, $X_{(2)}^i$ is fully determined by symmetry to be~\cite{Schwimmer:2008yh}
\be\label{eq-X2}
X_{(2)}^i = \frac{1}{2(d-2)} D^a\partial_aX^i_{(0)} = - \frac{1}{2(d-2)}K^i,
\ee
where $K^i$ denotes the extrinsic curvature of the $X_{(0)}^i$ surface, but we are leaving off the $(0)$ in our notation to save space.

The computation of $X_{(4)}^i$ is straightforward but tedious. We will only need to know $k_iX_{(4)}^i$ (where indices are being raised and lowered with $g^{(0)}_{ij}$), and the answer turns out to be
\begin{align}\label{eq:X4}
4(d-4)X_{(4)}^k &=2X_{(2)}^k\left(P^{jm}g_{jm}^{(2)}-4(X_{(2)})^2\right) \nonumber\\
&+K_{ab}^kg^{ab}_{(2)}+ 4g^{(2)}_{km} X_{(2)}^m+2X^{(2)}_jK^j_{ab}K^{kab} + k_iD_aD^a X^i_{(2)} \nonumber\\
&+ k^j (\nabla_n g^{(2)}_{jm}-\frac{1}{2}\nabla_j g^{(2)}_{mn})P^{mn} + X_{(2)}^n R_{kmnj}P^{jm}\nonumber\\ 
&+8\kappa\sigma_{(k)}^{ab}C_{kalb}-2(d-4)\Gamma^k_{jm}X^j_{(2)}X_{(2)}^m.
\end{align}
Here $\kappa$ depends on $\lambda_{\rm GB}$ as in \eqref{eq-kappa}. Notice that the last term in this expression is the only source of noncovariant-ness. One can confirm that this noncovariant piece is required from the definition of $X_{(4)}^i$---despite its index, $X_{(4)}^i$ does not transform like a vector under boundary diffeomorphisms.

We also note that the terms in $X^k_{(4)}$ with covariant derivatives of $g^{(2)}_{ij}$ can be simplified using the extended $k^i$ and $l^i$ fields described \S\ref{sec-kl} and the Bianchi identity: 
\begin{align}
k^j (\nabla_n g^{(2)}_{jm}-\frac{1}{2}\nabla_j g^{(2)}_{mn})P^{mn} &= -\frac{1}{4(d-1)}\nabla_kR +\frac{1}{d-2} \nabla_l R_{kk}.
\end{align}

Finally, we record here the formula for $X_{(4,{\rm log})}^k$ which is obtained from $X_{(4)}^k$ by multiplying by $4-d$ and sending $d\to 4$:
\begin{align}\label{eq:X4log}
-4X_{(4, {\rm log})}^k &=2X_{(2)}^k\left(P^{jm}g_{jm}^{(2)}-4(X_{(2)})^2\right) \nonumber\\
&+K_{ab}^kg^{ab}_{(2)}+ 4g^{(2)}_{km} X_{(2)}^m+2X^{(2)}_jK^j_{ab}K^{kab} + k_iD_aD^a X^i_{(2)} \nonumber\\
&+ k^j (\nabla_n g^{(2)}_{jm}-\frac{1}{2}\nabla_j g^{(2)}_{mn})P^{mn} + X_{(2)}^n R_{kmnj}P^{jm}\nonumber\\ 
&+8\kappa\sigma_{(k)}^{ab}C_{kalb}.
\end{align}
We will not bother unpacking all of the definitions, but the main things to notice is that the noncovariant part disappears.


\section{Details of the EWN Calculations}\label{EWN}
In this section we provide some insight into the algebra necessary to complete the calculations of the main text, primarily regarding the calculation of the subleading part of $(\delta \bar{X})^2$ in \S\ref{state-ind}. The task is to simplify \eqref{eq-subleadinginitial},
\begin{align}
\left.L^{-2}(\delta \bar{X})^2\right|_{z^2} &= 2k_i \delta X^i_{(4)} + 2g_{ij}^{(2)}k^i\delta X_{(2)}^j+g_{ij}\delta X_{(2)}^i\delta X_{(2)}^j+g_{ij}^{(4)}k^ik^j+X^m_{(4)}\partial_mg_{ij}k^ik^j\nonumber\\
&+2X_{(2)}^m\partial_mg_{ij}k^i\delta X_{(2)}^j+X^m_{(2)}\partial_mg_{ij}^{(2)}k^ik^j+\frac{1}{2}X_{(2)}^mX_{(2)}^n\partial_m\partial_ng_{ij}k^ik^j.
\end{align}
After some algebra, we can write this as
\be
\left.L^{-2}(\delta \bar{X})^2\right|_{z^2} = g^{(4)}_{kk}+ 2 \delta(X_{(4,{\rm cov})}^k) +2 g^{(2)}_{ik} \nabla_k X_{(2)}^i + \nabla_k X^{(2)}_j  \nabla_k X_{(2)}^j-\frac{1}{d-2}(X_{(2)}^l)\nabla_k R_{kk}.
\ee
Here we have defined
\be
X_{(4,{\rm cov})}^i = X_{(4)}^i + \frac{1}{2}\Gamma^i_{lm} X_{(2)}^l X_{(2)}^m,
\ee
which transforms like a vector (unlike $X_{(4)}^i$). From here, the algebra leading to \eqref{EWN Subleading} is mostly straightforward, though tedious. The two main tasks which require further explanation are the simplification of one of the terms in $g^{(4)}_{kk}$ and one of the terms in $\delta X_{(4,{\rm cov})}^{k}$. We will explain those now.

\paragraph{$g^{(4)}_{kk}$ Simplification}

We recall the formula for $g_{kk}^{(4)}$ from \eqref{eq-g4kk}:
\begin{align}
g^{(4)}_{kk} &= \frac{1}{d-4}\left[\kappa C_{kijm}C_k^{~ijm}+\frac{1}{ 8(d-1)}\nabla_k^2R - \frac{1}{4(d-2)}k^ik^j  \square R_{ij} \right.\nonumber\\
&\left.-\frac{1}{2(d-2)}R^{ij}R_{kikj}+\frac{d-4}{2(d-2)^2}R_{ki}R_k^i + \frac{1}{(d-1)(d-2)^2}RR_{kk} \right].
\end{align}
The main difficulty is with the term $k^ik^j\square R_{ij}$. We will rewrite this term by making use of the geometric quantities introduced in the other appendices, and in particular we make use of the extended $k$ and $l$ field from \S\ref{sec-kl}. We first separate it into two terms:
\be
k^ik^j  \square R_{ij} =  k^ik^j N^{rs} \nabla_r\nabla_s R_{ij} + k^ik^j P^{rs} \nabla_r\nabla_s R_{ij}.
\ee
Now we compute each of these terms individually:
\begin{align}
\begin{split}
k^ik^j N^{rs} \nabla_r\nabla_s R_{ij} &=  2k^ik^j l^s \nabla_k\nabla_s R_{ij} +2R_{kmlk}R^m_k\\ 
&=  2 \nabla_k\nabla_l R_{kk}  + 2w^ck^ik^j D_c R_{ij}+2R_{kmlk}R^m_k\\ 
&=  2 \nabla_k\nabla_l R_{kk} + 2w^cD_c R_{kk} - 4w^cw_cR_{kk}- 4w^cK_{ck}^aR_{ka}+2R_{kmlk}R^m_k\\
&=  2 \nabla_k\nabla_l R_{kk} + 2w^cD_c R_{kk} - 4w^cw_cR_{kk}+2R_{kmlk}R^m_k.
\end{split}
\end{align}
In the last line we assumed that $\sigma_{(k)}=0$ and $\theta_{(k)} =0$, which is the only case we will need to worry about. The other term is slightly messier, becoming
\begin{align}
\begin{split}
k^ik^j P^{rs} \nabla_r\nabla_s R_{ij} &=k^ik^j e^{sc} D_c\nabla_s R_{ij} \\
&=D_c(k^ik^j D^c R_{ij} ) - D_c(k^ik^j e^{sc}) \nabla_s R_{ij}\\ 
&=D_c(k^ik^j D^c R_{ij} ) - 2w_cD^cR_{kk}+4w_cw^cR_{kk}+6w_cK^{ca}_kR_{ak}\\
&~~~~ - 2K^{ca}_kD_cR_{ka}+ 2K^{ca}_k K_{ca}^iR_{ik}+ 2K^{ca}_k K_{c}^{bk}R_{ab}+ K^s \nabla_s R_{kk}\\ 
&=D_cD^c R_{kk}-2D_c(w^cR_{kk})-2D_c( K^{cak}R_{ka} ) - 2w_cD^cR_{kk}+4w_cw^cR_{kk}+6w_cK^{ca}_kR_{ak}\\
&~~~~ - 2K^{ca}_kD_cR_{ka}+ 2K^{ca}_k K_{ca}^iR_{ik}+ 2K^{ca}_k K_{c}^{bk}R_{ab}+ K^s \nabla_s R_{kk}\\ 
&=D_cD^c R_{kk}-2D_c(w^cR_{kk})-2D_c( K^{cak})R_{ka}  - 2w_cD^cR_{kk}+4w_cw^cR_{kk}+ K^s \nabla_s R_{kk}.
\end{split}
\end{align}
In the last line we again assumed that $\sigma_{(k)}=0$ and $\theta_{(k)} =0$. Putting the two terms together leads to some canellations:
\begin{align}\label{eq:boxR}
\begin{split}
k^ik^j  \square R_{ij} &=2 \nabla_k\nabla_l R_{kk}+2R_{kmlk}R^m_k + D_cD^c R_{kk}-2D_c(w^cR_{kk})\\
&-2(D_a\theta_{(k)} + R_{kcac})R^a_{k}+ K^s \nabla_s R_{kk}.
\end{split}
\end{align}

\paragraph{$\delta X_{(4,{\rm cov})}^{k}$ Simplification}
The most difficult term in \eqref{eq:X4}, which also gives the most interesting results, is
\be
k_iD_aD^a X^i_{(2)} = -\frac{1}{2(d-2)} (D_a-w_a)^2\theta_{(k)}+\frac{1}{2(d-2)} K_{ab}K^{abi} K_i.
\ee
The interesting part here is the first term, so we will take the rest of this section to discuss its variation. The underlying formula is \eqref{eq-wvar},
\be
\delta w_a =-w^cK_{ac} +R_{klak}.
\ee
From this we can compute the following related variations, assuming that $\theta_{(k)} =0$ and $\sigma_{(k)}=0$:
\begin{align}
\delta(D^aw_a) &= D^aR_{klak} + w^a\partial_a\theta_{(k)}-3D_a(K^{ab}_kw_b)\\
\delta(w^aD_a\theta_{(k)}) &= -3K_k^{ab}w_aD_b\theta_{(k)} +R_{klak}D^a\theta_{(k)} + w^aD_a\dot\theta_{(k)}\\
\delta(D^aD_a\theta_{(k)}) &= D^aD_a \dot\theta- \partial_a\theta_{(k)}\partial^a\theta_{(k)}-2P^{jm}R_{kjbm}D^b\theta_{(k)}.
\end{align}
Here $\dot\theta_{(k)} \equiv \delta \theta_{(k)}$ is given by the Raychaudhuri equation. We can combine these equations to get
\begin{align}
\delta\left( (D_a-w_a)^2 \theta_{(k)}  \right)&= \delta\left(D^aD_a\theta_{(k)}  \right)-2\delta\left( w^aD_a\theta_{(k)} \right)-\delta\left( (D_aw^a)\theta_{(k)}  \right)+\delta\left( w_aw^a \theta_{(k)}  \right)\nonumber\\
&= -D^aD_aR_{kk}+2 w^aD_aR_{kk}+(D_aw^a)R_{kk} -w_aw^aR_{kk}\nonumber\\
&-\frac{d}{d-2}(D_a \theta_{(k)})^2 -2R_{kb}D^b\theta_{(k)}-2(D\sigma)^2.
\end{align}

\section{The $d=4$ Case}\label{d=4}

As mentioned in the main text, many of our calculations are more complicated in even dimensions, though most of the end results are the same. The only nontrivial even dimension we study is $d=4$, so in this section we record the formulas and special derivations necessary for understanding the $d=4$ case. Some of these have been mentioned elsewhere already, but we repeat them here so that they are all in the same place.

\paragraph{Log Terms}

In $d=4$ we get log terms in the extremal surface, the metric, and the EWN inequality. By looking at the structure of the extremal surface equation, it's easy to see that the log term in in the extremal surface is related to $X^i_{(4)}$ in $d\neq 4$ by first multipling by $4-d$ and then setting $d\to 4$. The result was recorded in \eqref{eq:X4log}, and we repeat it here:
\begin{align}
-4X_{(4, {\rm log})}^k &=2X_{(2)}^k\left(P^{jm}g_{jm}^{(2)}-4(X_{(2)})^2\right) \nonumber\\
&+K_{ab}^kg^{ab}_{(2)}+ 4g^{(2)}_{km} X_{(2)}^m+2X^{(2)}_jK^j_{ab}K^{kab} + k_iD_aD^a X^i_{(2)} \nonumber\\
&+ k^j (\nabla_n g^{(2)}_{jm}-\frac{1}{2}\nabla_j g^{(2)}_{mn})P^{mn} + X_{(2)}^n R_{kmnj}P^{jm}\nonumber\\ 
&+8\kappa\sigma_{(k)}^{ab}C_{kalb}.
\end{align}
There is a similar story for $g_{kk}^{(4,{\rm log})}$, which was recorded earlier in \eqref{eq-glog}:
\begin{align}
g^{(4, {\rm log})}_{kk} &=-\left[\kappa C_{kijm}C_k^{~ijm}+\frac{1}{ 24}\nabla_k^2R - \frac{1}{8}k^ik^j  \square R_{ij} -\frac{1}{4}R^{ij}R_{kikj} + \frac{1}{12}RR_{kk} \right].
\end{align}
From these two equations, it is easy to see that the log term in $(\delta \bar{X})^2$ has precisely the same form as the subleading EWN inequality \eqref{EWN Subleading} in $d\geq 5$, except we first multiply by $4-d$ and then set $d\to 4$. This results in
\begin{align}
\left.L^{-2}(\delta \bar{X})^2\right|_{z^2\log z, d=4}  &=-\frac{1}{4}(D_a\theta_{(k)} + R_{ka})^2-\frac{1}{4}(D_a\sigma^{(k)}_{bc})^2.
\end{align}
Note that the Gauss-Bonnet term drops out completely due to special identities of the Weyl tensor valid in $d=4$~\cite{Fu:2017aa}. The overall minus sign is important because $\log z$ should be regarded as negative.

\paragraph{QNEC in Einstein Gravity}

For simplicity we will only discuss the case of Einstein gravity for the QNEC in $d=4$, so that the entropy functional is just given by the extremal surface area divided by $4G_N$. At order $z^2$, the norm of $\delta \bar{X}^\mu$ is formally the same as the expression in other dimensions:
\begin{align}
\left. L^{-2} (\delta \bar{X})^2\right|_{z^2} &=g^{(4)}_{kk} + 2g_{ik}^{(2)}\nabla_kX_{(2)}^i + \nabla_kX_j^{(2)}\nabla_kX_{(2)}^j - \frac{1}{2}X_{(2)}^l \nabla_kR_{kk}+ 2\delta (k_i X^i_{(4){\rm cov}}).
\end{align}
Now, though, $X_{(4)}^k$ and $g_{kk}^{(4)}$ are state-dependent and must be related to the entropy and energy-momentum, respectively.

We begin with the entropy. From the calculus of variations, we know that the variation of the extremal surface area is given by
\be
\delta A = -\lim_{\epsilon\to 0}\frac{L^3}{\epsilon^{3}}\int \sqrt{h}\frac{1}{\sqrt{1+g_{nm}\partial_z\bar{X}^{n}\partial_z\bar{X}^m}} g_{ij}\partial_z\bar{X}^i \delta X^j .
\ee
A few words about this formula are required. The $\bar{X}^\mu$ factors appearing here must be expanded in $\epsilon$, but the terms without any $(n)$ in their notation do {\em not} refer to ${(0)}$, unlike elsewhere in this paper. The reason is that we have to do holographic renormalization carefully at this stage, and that means the boundary conditions are set at $z=\epsilon$. So when we expand out $\bar{X}^\mu$ we will find its coefficients determined by the usual formulas in terms of $X_{(0)}^i$. We need to then solve for $X_{(0)}^i$ in term of $X^i \equiv \bar{X}^i(z=\epsilon)$ re-express the result in terms of $X^i$ alone. Since we are not in a high dimension this task is relatively easy. An intermediate result is
\begin{align}
\left. \frac{k^i}{L^3\sqrt{h}} \frac{\delta A}{\delta X^i} \right|_{\epsilon^0} &= - 2\left.X_{(2)}^k\right|_{\epsilon^2} -4\left(X_{(4)}^k -(X_{(2)})^2X_{(2)}^k\right) - X_{(4,{\rm log})}^k.
\end{align}
The notation on the first term refers to the order $\epsilon^2$ part of $X_{(2)}^i$ that is generated when $X_{(2)}^i$ is written in terms of $\bar{X}^i(z=\epsilon)$. The result of that calculation is
\begin{align}
-4\left.X_{(2)}^k\right|_{\epsilon^2} &= 2X^{(2)}_jK^{jab}K_{ab}^ik_i +k_iD^bD_b X_{(2)}^i +K^m\Gamma_{ml}^i X_{(2)}^lk_i \nonumber\\
&+g_{(2)}^{ab}K_{ab}^ik_i+P^{kj}R^i_{jmk}  X_{(2)}^m k_i+ k^m\left(\nabla_jg^{(2)}_{mk} -\frac{1}{2}\nabla_mg^{(2)}_{jk}\right) P^{jk} \nonumber\\
&=-4X_{(4,{\rm log})}^k  - 2X_{(2)}^k\left(P^{jm}g_{jm}^{(2)}-4(X_{(2)})^2\right) -4g^{(2)}_{km} X_{(2)}^m +K^m\Gamma_{ml}^i X_{(2)}^lk_i  .
\end{align}
We have dropped terms of higher order in $\epsilon$. Thus we can write
\begin{align}
\left. \frac{k^i}{L^3\sqrt{h}} \frac{\delta A}{\delta X^i} \right|_{\epsilon^0} &= -3X_{({\rm log})}^k  - X_{(2)}^kP^{jm}g_{jm}^{(2)}+8X_{(2)}^k(X_{(2)})^2 -2g^{(2)}_{km} X_{(2)}^m  -4X_{(4){\rm cov}}^k.
\end{align}
We will want to take one more variation of this formula so that we can extract $\delta X^k_{(4){\rm cov}}$. We can get some help by demanding that the $z^2\log z$ part of EWN be saturated, which states
\be
g^{(\log)}_{kk} + 2\delta X_{\rm log}^k = 0.
\ee
Then we have
\begin{align}
\delta\left( \left. \frac{k^i}{L^3\sqrt{h}} \frac{\delta A}{\delta X^i} \right|_{\epsilon^0} \right)&= \frac{3}{2}g^{(\log)}_{kk}  - \delta(X_{(2)}^kP^{jm}g_{jm}^{(2)})+8\delta(X_{(2)}^k(X_{(2)})^2) -2\delta(g^{(2)}_{km} X_{(2)}^m)  -4\delta X_{(4){\rm cov}}^k.
\end{align}
Assuming that $\theta_{(k)} = \sigma_{(k)} =0$, we can simplify this to
\begin{align}
\delta\left( \left. \frac{k^i}{L^3\sqrt{h}} \frac{\delta A}{\delta X^i} \right|_{\epsilon^0} \right)&= \frac{3}{2}g^{(\log)}_{kk}  -\frac{1}{4} R_{kk} P^{jm}g_{jm}^{(2)} -\frac{1}{4}\nabla_k(\theta_{(l)} R_{kk})-\frac{1}{2}g^{(2)}_{kl}R_{kk}  -4\delta X_{(4){\rm cov}}^k.
\end{align}
We can combine this with the holographic renormalization formula~\cite{deHaro:2000vlm}
\begin{align}
g^{(4)}_{kk} &= 4\pi G_N L^{-3}T_{kk} + \frac{1}{2}(g_{(2)}^2)_{kk} -\frac{1}{4}g^{(2)}_{kk} g^{ij}g^{(2)}_{ij}-\frac{3}{4}g^{(\rm log)}_{kk}\nonumber\\
&= 4\pi G_N L^{-3}T_{kk} + \frac{1}{8}R_{k}^i R_{ik} -\frac{1}{16}R_{kk}R-\frac{3}{4}g^{(\rm log)}_{kk}
\end{align}
to get
\begin{align}
\left. L^{-2}(\delta \bar{X}^i)^2\right|_{z^2} &= 4\pi G_NL^{-3} T_{kk}-\frac{1}{2}\delta\left( \left. \frac{k^i}{L^3\sqrt{h}} \frac{\delta A}{\delta X^i} \right|_{\epsilon^0} \right).
\end{align}
After dividing by $4G_N$, we recognize the QNEC.


\bibliographystyle{utcaps}
\bibliography{all}

\end{document}